\documentclass[11pt, a4paper]{article}
\usepackage{times}
\usepackage[T1]{fontenc}
\normalfont

\usepackage{amsmath,amssymb}
\usepackage{bm}
\usepackage[pdftex]{graphicx}% Include figure files
\usepackage{wrapfig}
\usepackage{cite}
\begin{document}
\baselineskip 0.6cm
%%%%% Title Page %%%%%
\begin{titlepage}
\begin{center}
%% Preprint Number
\begin{flushright}
EPHOU-19-013
\end{flushright}
\vskip 1cm
%% Title
{\Large \bf 
Modular $A_4$ invariance and leptogenesis
}\\
\vskip 1.5cm
%% Authors
{\large 
Takehiko Asaka$^{1}$,
Yongtae Heo$^{2}$,
Takuya H. Tatsuishi$^{3}$, and 
Takahiro Yoshida$^{2}$\\
}
\vskip 0.5cm
%% Addresses
$^1${\em
Department of Physics, Niigata University, Niigata 950-2181, Japan
}

$^2${\em
Graduate School of Science and Technology, Niigata University, Niigata 950-2181, Japan
}

$^3${\em
Department of Physics, Hokkaido University, Sapporo 060-0810, Japan
}
\vskip 0.3cm
%% Date
(September 14, 2019)
\vskip 2cm
%% Abstract
\begin{abstract}
We consider a model with three right-handed neutrinos in which 
Yukawa coupling constants and Majorana masses are obtained by 
requiring the modular $A_4$ symmetry. 
It has been shown that the model can explain
mass hierarchies and mixing patterns of charged leptons and neutrinos
with the seesaw mechanism.  
In this article we investigate the leptogenesis by decays of right-handed neutrinos
in this model.  
It is shown that masses of right-handed neutrinos 
are about $10^{13}$ GeV in order to account for the observed baryon asymmetry of the universe.
Furthermore, the positive sign of the baryon asymmetry is obtained only for 
the limited ranges of mixing angles and CP violation phases of active neutrinos,
which can be tested by future neutrino experiments.
\end{abstract}

\end{center}
\end{titlepage}

\title{
\begin{flushright}
\ \\*[-80pt] 
\begin{minipage}{0.2\linewidth}
\normalsize
%arXiv:YYMM.NNNN \\
%XXX-YYY-ZZZ \\
EPHOU-19-XXX \\*[50pt]
\end{minipage}
\end{flushright}
{\Large \bf 
 Modular $A_4$ invariance and leptogenesis
\\*[20pt]}}

%%%%%%%%%%%%%%%%%%%%%%%%%%%%%%%%%%%%%%%%%%%%%%%%%%%%%%%%%%%%%%%%%%%%%%%%%%%%%%%%%%%%   Introduction   %%%%%%%%%%%%%%%%%%%%%%%%%%
%%%%%%%%%%%%%%%%%%%%%%%%%%%%%%%%%%%%%%%%%%%%%%%%%%%%%%%%%%%%%%%%%%%%%%%%%%%%%
\section{Introduction}
The standard model (SM) is well established by the discovery of the Higgs boson.
There are, however, various unsolved problems, and the flavor puzzle is one of those problems.
One interesting approach to the origin of flavor structure is to impose a flavor symmetry on a theory.
Especially, to explain the large mixing angles in the lepton sector, lepton flavor models with non-Abelian discrete symmetry such as $S_3$, $A_4$, $S_4$, $A_5$, and other groups have been studied \cite{Altarelli:2010gt,Ishimori:2010au,Ishimori:2012zz,King:2013eh,King:2014nza}.

Recently, lepton flavor model with $A_4$ being a subgroup of modular group has been proposed \cite{Feruglio:2017spp}.
Modular symmetry is a geometrical symmetry of a two-dimensional torus $T^2$, and the modulus is a complex field deciding the shape of the torus. 
Modular group induces $S_3$, $A_4$, $S_4$, or $A_5$ as the finite quotient group \cite{deAdelhartToorop:2011re}.
In this framework, Yukawa couplings are written in terms of modular forms, which are non-trivial representation of flavor symmetry and constrained as explicit functions of the modulus.
These features of Yukawa couplings enable us to construct flavor models without flavons.
The lepton models with
$S_3$ \cite{Kobayashi:2018vbk},
$A_4$ \cite{Feruglio:2017spp,Kobayashi:2018vbk,Criado:2018thu,Kobayashi:2018scp,Novichkov:2018yse,Nomura:2019jxj,Kobayashi:2019mna,Ding:2019zxk,Kobayashi:2019xvz}, 
$S_4$ \cite{Penedo:2018nmg,Novichkov:2018ovf,King:2019vhv}, and
$A_5$ \cite{Novichkov:2018nkm,Ding:2019xna}
have been studied.
Moreover,
quark model \cite{Okada:2018yrn},
combination of lepton and quark models \cite{Kobayashi:2018wkl}, and
GUT model \cite{deAnda:2018ecu,Kobayashi:2019rzp}
have also been studied.

Modular symmetry is also interesting in the viewpoint of superstring theory.
The torus compactification is a simple compatification of the extra dimensions, and which leads modular symmetry as explained above.
Moreover, the orbifold compactification as well as magnetized torus compatification leads flavor symmetry including modular group or its finite subgroups \cite{Cremades:2004wa,Kobayashi:2017dyu,Kobayashi:2018rad,Kobayashi:2018bff,Baur:2019kwi,Kariyazono:2019ehj}.
In this sense, the modular symmetry or its finite subgroups can be expected as geometrical symmetries of extra dimensions.

Interestingly,  models with the modular symmetry can predict the patterns of masses and mixing angles of charged leptons and neutrinos by using a very limited number of parameters. It has been discussed that the neutrino masses are generated by introducing the Weinberg's dimension five operators or right-handed neutrinos with the seesaw mechanism, and both possibilities have been shown to be successful. It is a natural question to investigate whether right-handed neutrinos in the models can also explain the baryon asymmetry of the universe (BAU) through the leptogenesis.

The BAU is now measured very precisely by the cosmic microwave background radiation as \cite{Aghanim:2018eyx}
\begin{align}
Y_B = \frac{n_B}{s} = ( 0.852-0.888 ) \times 10^{-10} \,,
\end{align}
where the BAU at the present universe, $Y_B$, is defined by the ratio between the number density of baryon asymmetry $n_B$ and the entropy density $s$.  This asymmetry should be generated before the beginning of the big-bang nucleosynthesis after the primordial inflation ends. One of the most studied scenarios for baryogensis is the canonical leptogenesis scenario \cite{Fukugita:1986hr} in which the decays of right-handed neutrinos can generate the lepton asymmetry that is partially converted into the baryon asymmetry \cite{Kuzmin:1985mm} via the sphaleron process \cite{Klinkhamer:1984di}. The sign and magnitude of the BAU are predicted by the masses and Yukawa coupling constants of right-handed neutrinos. 

The absolute mass scales of right-handed neutrinos cannot be determined by the data of the neutrino oscillations and the BAU. When their masses are hierarchical,  the lightest one must be larger than ${\cal O}(10^9)$~GeV~\cite{Davidson:2002qv,Giudice:2003jh} to explain the BAU.  It can be, however, small as TeV scale if right-handed neutrinos are quasi-degenerate in mass~\cite{Pilaftsis:2003gt}. 

Furthermore, the sign of the BAU is controlled by the CP violation pattern in leptonic sector.   Note that the sign of the BAU cannot be predicted uniquely even if the CP violations associated with active neutrinos (i.e., the Dirac and Majorana phases in the mixing matrix of active neutrinos) are determined. This is because there exist one or more additional phases associated with right-handed neutrinos which decouple from the low energy phenomena if they are sufficiently heavy. 
Under these situations, it is interesting to investigate the sign and magnitude of the BAU in the models with the modular symmetry, since there are non-trivial relations between the properties of right-handed neutrinos and the low energy observables of neutrino physics due to the symmetry. As our first work, we shall discuss the leptogenesis in the model with $A_4$ symmetry~\cite{Kobayashi:2018scp} simply because the model has a small number of free parameters and then very predictive.

The paper is organized as follows. In section 2, we briefly review the modular symmetry in the framework of the theory with extra dimensions which is compactified on a torus. We then explain the model with the $A_4$ symmetry in section 3. The leptogenesis in the model is discussed in section 4. We present in section 5 the results of the analysis, namely the sign and magnitude of the BAU predicted by the model. The final section is devoted to conclusions.

%%%%%%%%%%%%%%%%%%%%%%%%%%%%%%%%%%%%%%%%%%%%%%%%%%%%%%%%%%%%%%%%%%%%%%%%%%%%%%%%

\vskip 1 cm
%%%%%%%%%%%%%%%%%%%%%%%%%%%%%%%%%%%%%%%%%%%%%%%%%%%%%%%%%%%%%%%%%%%%%%
%%%%%%%%%%%%%%%%%%%%%%%%%%%%%%%%%%%%%%%%%%%%%%%%%%%%%%%%%%%%%%%%%%%%%%
%%%%%%%%%%%%%%%%%%%%%%%%%%%%%%%%%%%%%%%%%%%%%%%%%%%%%%%%%%%%%%%%%%%%%%
\section{Modular group and its finite quotient subgroups}

In this section, we give a brief review on the modular symmetry on a torus.
A two-dimensional torus $T^2$ can be constructed by $\mathbb{R}^2/\Lambda$, where $\Lambda$ denotes a two-dimensional lattice.
We use the complex coordinate on $\mathbb{R}^2$ and denote basis vectors of $\Lambda$ as $\alpha_1=2\pi R$ and $\alpha_2=2\pi R\tau$, where $R$ is real and $\tau$ is a modulus belonging to upper-half complex plane $\textrm{Im}\,\tau>0$.
There is some ambiguity in choice of the basis vectors.
The same lattice can be spanned by the following basis vectors,
\begin{equation}
	\begin{pmatrix} \alpha_2^\prime \\ \alpha_1^\prime \\ \end{pmatrix}
=\begin{pmatrix} a & b \\ c & d \\ \end{pmatrix}
\begin{pmatrix} \alpha_2 \\ \alpha_1 \\ \end{pmatrix},\quad
\begin{pmatrix} a & b \\ c & d \\ \end{pmatrix}\in SL(2,\mathbb{Z}),
\end{equation}
where
\begin{equation}
	SL(2,\mathbb{Z})=\left\{\left.\begin{pmatrix} a & b \\ c & d \\ \end{pmatrix}\right| a,b,c,d\in\mathbb{Z},~~ ad-bc=1 \right\}\equiv\Gamma.
\end{equation}
This transformation of basis vectors is written in terms of the modulus $\tau\equiv\alpha_2/\alpha_1$ by
\begin{equation}
	\tau \to \tau^\prime = \gamma\tau = \frac{a\tau+b}{c\tau+d},\quad
\begin{pmatrix} a & b \\ c & d \\ \end{pmatrix}\in SL(2,\mathbb{Z}).
\label{eq:modular transf_tau}
\end{equation}
The modular group is the transformation group acts on the modulus preserving the lattice $\Lambda$.
Since $\gamma$ and $-\gamma$ transform $\tau$ in the same way in (\ref{eq:modular transf_tau}), the modular group is isomorphic to $SL(2,\mathbb{Z})/\{\mathbb{I},-\mathbb{I}\}\equiv\overline{\Gamma}$.
The modular group is generated by two generators $S$ and $T$,
\begin{equation}
	S=\begin{pmatrix} 0 & 1 \\ -1 & 0 \\ \end{pmatrix},\quad
T=\begin{pmatrix} 1 & 1 \\ 0 & 1 \\ \end{pmatrix}
\end{equation}
In terms of the modulus, they induce the transformations,
$S:\tau \to -1/\tau$ and $T: \tau \to \tau+1$.
We can easily see that they satisfy the following algebraic relations,
$S^2 = \mathbb{I}$ and $(ST)^3=\mathbb{I}$.
We introduce a series of groups $\Gamma(N)$, $N=1,2,3,\dots$ called principal congruence subgroups,
\begin{equation}
	\Gamma(N)=\left\{\begin{pmatrix} a & b \\ c & d \\ \end{pmatrix}\in SL(2,\mathbb{Z}),~
\begin{pmatrix} a & b \\ c & d \\ \end{pmatrix}=\begin{pmatrix} 1 & 0 \\ 0 & 1 \\ \end{pmatrix} (\textrm{mod}\,N)\right\}.
\end{equation}
We also define $\overline{\Gamma}(N)=\Gamma(N)/\{\mathbb{I},-\mathbb{I}\}$ for $N=1,2$ and $\overline{\Gamma}(N)=\Gamma(N)$ for $N>2$.
The groups $\overline{\Gamma}(N)$ are infinite subgroups of the modular group.
The quotient groups defined $\Gamma_N\equiv\overline{\Gamma}/\overline{\Gamma}(N)$ are finite subgroups of the modular group, called finite modular groups.
In the finite modular groups $\Gamma_N$, generators obey additional, algebraic relation $T^N=\mathbb{I}$.
The groups $\Gamma_N$ with $N=2,3,4,5$ are isomorphic to $S_3$, $A_4$, $S_4$, and $A_5$, respectively \cite{deAdelhartToorop:2011re}.

Modular forms $f(\tau)$ of weight $k$ and level $N$ are holomorphic functions transforming under the $\Gamma(N)$ as
\begin{equation}
	f(\gamma\tau)=(c\tau+d)^kf(\tau),\quad\gamma\in\Gamma(N),
\end{equation}
where $k$ is even and non-negative value and called modular weight.
In the case of $\Gamma_3\simeq A_4$, the explicit form of $A_4$ triplet modular forms of weight 2, $Y^{A_4} (\tau) = (Y_1 (\tau) , Y_2 (\tau) , Y_3(\tau))$, is obtained as~\cite{Feruglio:2017spp} 
\begin{align} 
\label{eq:Y-A4}
	Y_1(\tau) &= \frac{i}{2\pi}\Big( \frac{\eta'(\tau/3)}{\eta(\tau/3)}  
	+\frac{\eta'((\tau +1)/3)}{\eta((\tau+1)/3)}  
	+\frac{\eta'((\tau +2)/3)}{\eta((\tau+2)/3)} 
	- \frac{27\eta'(3\tau)}{\eta(3\tau)}  \Big) \,, 
	\nonumber \\
	Y_2(\tau) &= \frac{-i}{\pi}\Big( \frac{\eta'(\tau/3)}{\eta(\tau/3)}  
	+\omega^2\frac{\eta'((\tau +1)/3)}{\eta((\tau+1)/3)}  
  +\omega \frac{\eta'((\tau +2)/3)}{\eta((\tau+2)/3)}  \Big) \,, 
  \\
	Y_3(\tau) &= \frac{-i}{\pi}\Big( \frac{\eta'(\tau/3)}{\eta(\tau/3)}  
	+\omega\frac{\eta'((\tau +1)/3)}{\eta((\tau+1)/3)}  
  +\omega^2 \frac{\eta'((\tau +2)/3)}{\eta((\tau+2)/3)}  \Big) 
  \,. \nonumber
\end{align}
where the Dedekind eta-function $\eta(\tau)$ is given by 
$\eta(\tau) = q^{1/24} \prod_{n =1}^\infty (1-q^n), q = e^{2 \pi i \tau}.$
%The $\eta(\tau)$ function behaves under $S$ and $T$ transformations as 
%\begin{equation}\label{eq:eta-ST}
%	\eta(-1/\tau) = \sqrt{-i \tau}\eta(\tau), \qquad \eta(\tau + 1) = e^{i\pi/12}\eta(\tau),
%\end{equation}
%and thus $\eta(\tau)^{24}$ is a modular form of weight $12$.
%$A_4$ triplet modular forms of wight 2 \cite{Feruglio:2017spp} are 
%given by  where

%and its $q$-expansions are written by
%\begin{eqnarray} 
%	Y_1(\tau) &=& 1 + 12 q + 36 q^2 + 12 q^3 + \cdots,  \nonumber \\
%	Y_2(\tau) &=& -6q^{1/3}(1 + 7 q + 8q^2 + \cdots) , \\
%	Y_3(\tau) &=& -18q^{2/3}(1 + 2 q + 5q^2 + \cdots) . \nonumber
%\end{eqnarray}

%Superstring theory on the torus $T^2$ or orbifold $T^2/Z_N$ has the modular symmetry.
%Its low-energy effective field theory is described in terms of supergravity theory,
%and the string-derived supergravity theory has also the modular symmetry.
Under the modular transformation (\ref{eq:modular transf_tau}), chiral superfields $\phi^{(I)}$ transform as \cite{Ferrara:1989bc},
\begin{equation}
	\phi^{(I)} \to (c\tau+d)^{-k_I}\rho^{(I)}(\gamma)\phi^{(I)},
\end{equation} 
where $-k_I$ is the modular weight and $\rho^{(I)}(\gamma)$ denotes an unitary representation matrix of $\gamma\in\Gamma_N$.
%The kinetic terms of their scalar components are written by 
%\begin{equation}
%	\sum_I\frac{|\partial_\mu\phi^{(I)}|^2}{\langle-i\tau+i\bar{\tau}\rangle^{k_I}}
%\end{equation}
%which is invariant under the modular transformation.
%Here, we use the convention that the superfield and its scalar component are denoted by the same letter.
A coupling constant for the $n$-th order term between $\phi^{(I_1)},\cdots, \phi^{(I_n)}$ should be a modular form of weight $k_Y(n)$ and a representation of $\Gamma_N$ transformed as
\begin{equation}
	Y_{I_1,I_2,\dots,I_n}(\gamma\tau)=(c\tau+d)^{k_Y(n)}\rho(\gamma)Y_{I_1,I_2,\dots,I_n}(\tau),
\end{equation}
where $\rho(\gamma)$ is representation of $\gamma$ for the modular form, and a modular invariant superpotential $W$ is written by
\begin{equation}
	W=Y_{I_1,I_2,\dots,I_n}(\tau)\phi^{(I_1)}\phi^{(I_2)}\cdots\phi^{(I_n)}
\end{equation}
satisfying $k_Y(n)=\sum_nk_{I_n}$ and $\rho(\gamma)\prod_n\rho^{(I_n)}(\gamma)=\mathbb{I}$.

We study the model which field content is the same as the minimal supersymmetric standard model (MSSM) extended by right-handed neutrinos in the following sections.
The superpotential of our model has vanishing modular weight.
We note that Yukawa couplings as well as higher order couplings depend on modulus and can have non-vanishing modular weights.
The breaking scale of supersymmetry (SUSY) can be between $\mathcal{O}(1)$TeV and the compactification scale. 
Here we take the breaking scale is sufficiently high, namely it is much higher than the masses of right-handed neutrinos, for simplicity.
Then, the lepton flavor physics and the leptogenesis can be discussed without SUSY.\footnote{In our scenario we assume that SUSY is broken at Planck scale or close to that. In this condition the masses of SUSY particles including gravitino are around Planck scale. Thus, the stringent constraint on the reheating temperature of the inflation from the gravitino problem can be avoided.}
The modular symmetry is broken by the vacuum expectation value of $\tau$ at the compactification scale which is the Planck scale or slightly lower scale order.

%The superpotential should be also invariant under the modular symmetry.
%In other words, the superpotential should have vanishing modular weight in global supersymmetric models.
%On the other hand, the superpotential in supergravity should be invariant under the modular symmetry up to the K\"ahler transformation.
%We study the model which field content is the same as the minimal supersymmetric standard model (MSSM) extended by right-handed neutrinos in the following sections.
%The superpotential of our model has vanishing modular weight.
%We note that Yukawa couplings as well as higher order couplings depend on modulus and can have non-vanishing modular weights.
%The breaking scale of supersymmetry can be between $\mathcal{O}(1)$TeV and the compactification scale. 
%Here we take the breaking scale is sufficiently high, namely it is much higher than the masses of right-handed neutrinos, for simplicity.
%Then, the lepton flavor physics and the leptogenesis can be discussed without supersymmetry.
%The modular symmetry is broken by the vacuum expectation value of $\tau$ at the compactification scale which is the Planck scale or slightly lower scale order.

%%%%%%%%%%%%%%%%%%%%%%%%%%%%%%%%%%%%%%%%%%%%%%%%%%%%%%%%%%%%%%%%%%%%%%
%%%%%%%%%%%%%%%%%%%%%%%%%%%%%%%%%%%%%%%%%%%%%%%%%%%%%%%%%%%%%%%%%%%%%%
%%%%%%%%%%%%%%%%%%%%%%%%%%%%%%%%%%%%%%%%%%%%%%%%%%%%%%%%%%%%%%%%%%%%%%
\section{Lepton flavor model with modular $A_4$ symmetry}

The $A_4$ flavor models with flavon field have been discussed in the lepton sector \cite{Altarelli:2010gt,Ishimori:2010au,Ishimori:2012zz,King:2013eh,King:2014nza}.
On the other hand, a modular invariant flavor model with the $A_4$ symmetry can explain the large mixing angles of lepton flavors without flavons.
One of the authors (THT) has already obtained a successful result of the lepton sector in $A_4$ modular symmetry~\cite{Kobayashi:2018scp}.
In order to clarify the difference in the flavor structure of mass matrices between the quarks and leptons,
we briefly summarize the previous results of the lepton sector and add discussions of the feature of the lepton model.

It is supposed that the three left-handed lepton doublets $L_i$ are compiled in a triplet of $A_4$.
The three right-handed neutrinos $N_i^c$ are compiled in a triplet of $A_4$.
On the other hand, the Higgs doublets, $H_{u,d}$, are supposed to be singlets of $A_4$. 
The three right-handed charged leptons are assigned for three different singlets of $A_4$ as $(e_1^c,e_2^c,e_3^c)=(e^c,\mu^c,\tau^c)=(1,1'',1')$.
Therefore, there are three independent couplings $\alpha$, $\beta$
and $\gamma$,  in the superpotential of the charged lepton sector.
Those coupling constants can be adjusted to the observed charged lepton masses.
The assignments of representations and modular weights to the MSSM fields and right-handed neutrino superfields are presented in Table~\ref{tb:fields0}.

%%%%%%%%%%%%%%%%%%%%%%%%%%%%%%%%%%%%%%%%%%%%%%%%%%%%%
%%%%%%%%%%%%%%%%%%%%%%%%%%%%%%%%%%%%%%%%%%%%%%%%%%%%%   
\begin{table}[th]
	\centering
	\begin{tabular}{|c||c|c|c|c|c|} \hline 
		&$L$&$e^c,~\mu^c,~\tau^c$&$N^c$&$H_u$&$H_d$\\ \hline \hline 
		\rule[14pt]{0pt}{0pt}
		$SU(2)_L$&$2$&$1$&$1$&$2$&$2$\\
		$A_4$&$3$& $1$,\ $1''$,\ $1'$&$3$&$1$&$1$\\
		$-k_I$&$-1$&$-1 $&$-1$&0&0 \\ \hline
	\end{tabular}
	\caption{
		The charge assignment of $SU(2)_L$, $A_4$, and the modular weight $-k_I$.}
	\label{tb:fields0}
\end{table}
%%%%%%%%%%%%%%%%%%%%%%%%%%%%%%%%%%%%%%%%%%%%%%%%%%%%%
%%%%%%%%%%%%%%%%%%%%%%%%%%%%%%%%%%%%%%%%%%%%%%%%%%%%%   

The modular invariant mass terms of leptons are given as the following superpotentials:
\begin{align}
	W_e&=\alpha \, e^c H_d(LY^{A_4})_{\bf 1}
		+\beta \, \mu^c H_d(LY^{A_4})_{\bf 1'}
		+\gamma \, \tau^c H_d(LY^{A_4})_{\bf 1"}~,\label{charged} \\
	W_D&=g_1 \, \big(N^c H_u (L Y^{A_4})_{{\bf 3}\rm s} \big)_{\bf 1}
	    +g_2 \, \big(N^c H_u (L Y^{A_4})_{{\bf 3}\rm a} \big)_{\bf 1}
	~,  \label{Dirac}\\
	W_N&=\Lambda \, (N^c N^c Y^{A_4})_{\bf 1}~, \label{Majorana}
\end{align}
where sums of the modular weights vanish.
The parameters $\alpha$, $\beta$, $\gamma$, and $g_{1,2}$ 
are coupling constants, and $\Lambda$ is a mass parameter for the 
Majorana masses for right-handed neutrinos. Following Ref. \cite{Kobayashi:2018scp}, we take $g_1$ and $g_2$ as real and complex parameters, respectively.
\begin{align}
g_2=|g_2|e^{i \phi_g}.
\end{align}
The functions $Y^{A_4}(\tau)$ are $A_4$ triplet modular forms of weight $2$ which components are shown in Eq.~(\ref{eq:Y-A4}). As for the field contents and the basis for $A_4$ group, see discussions in Ref. \cite{Kobayashi:2018scp}.

%%%%%%%%%%%%%%%%%%%%%%%%%%%%%%%%%%%%%%%%%%%%%%%%%%%%%%%%%%%%%%%%%%%
%%%%%%%%%%%%% Charged lepton mass matrix  %%%%%%%%%%%%%%%%%%%%%%%%%
%%%%%%%%%%%%%%%%%%%%%%%%%%%%%%%%%%%%%%%%%%%%%%%%%%%%%%%%%%%%%%%%%%%
The superpotential (\ref{charged}) leads to the following charged leptons mass matrix:
\begin{align}
\begin{aligned}
	M_E&=v_d~ {\rm diag}[\alpha, \beta, \gamma]
	\begin{pmatrix}
	Y_1 & Y_3 & Y_2 \\
	Y_2 & Y_1 & Y_3 \\
	Y_3 & Y_2 & Y_1
	\end{pmatrix}_{RL},
\end{aligned}\label{eq:CL}
\end{align}
where  $v_d=\langle H_d \rangle $.
Note that we should evaluate the mass matrix  at the SUSY-breaking
scale and also include the corrections due to the RGE evolution and
SUSY-breaking. In this analysis we neglect these effects without
specifying the scale and the mediation mechanism of SUSY-breaking.
Such corrections have been discussed, for example in Ref.\cite{Criado:2018thu}.
%and we omit the superscript $A_4$ of $Y_i^{A_4}$ hereafter.
The coefficients $\alpha$, $\beta$, and $\gamma$ 
are taken to be real positive by rephasing right-handed charged lepton fields
without loss of generality.
Those parameters can be written in terms of the modulus $\tau$ and the charged lepton masses together with $v_d$.
%%%%%%%%%%%%%%%%%%%%%%%%%%%%%%%%%%%%%%%%%%%%%%%%%%%%%%%%%%
%%%%%%%%%%%%%%%%%  Neutrino mass matrix %%%%%%%%%%%%%%%%%%
%%%%%%%%%%%%%%%%%%%%%%%%%%%%%%%%%%%%%%%%%%%%%%%%%%%%%%%%%%
The superpotential (\ref{Dirac}) gives the Dirac neutrino mass matrix:
\begin{align}
	M_D=v_u\begin{pmatrix}
	2g_1Y_1 & (-g_1+g_2)Y_3 & (-g_1-g_2)Y_2 \\
	(-g_1-g_2)Y_3 & 2g_1Y_2 & (-g_1+g_2)Y_1 \\
	(-g_1+g_2)Y_2 & (-g_1-g_2)Y_1 & 2g_1Y_3\end{pmatrix}_{RL},
	\label{MD}
\end{align}
where $v_u=\langle H_u \rangle $.
On the other hand, the right-handed Majorana neutrino mass matrix is obtained from the superpotential~(\ref{Majorana}):
\begin{align}
	M_N=\Lambda\begin{pmatrix}
	2Y_1 & -Y_3 & -Y_2 \\
	-Y_3 & 2Y_2 & -Y_1 \\
	-Y_2 & -Y_1 & 2Y_3\end{pmatrix}_{RR}.
	\label{MajoranaR}
\end{align}
Finally, the effective neutrino mass matrix is obtained through the type I seesaw as follows:
\begin{align}
\label{eq:seesaw-mass}
	M_\nu=-M_D^{\rm T} M_N^{-1}M_D^{} ~.
\end{align}
The masses of active neutrinos, $m_i$, are found by diagonalizing 
$M_\nu$, and the lepton mixing matrix $U$ in the charged current 
is also found by the diagonalization of $M_\nu$ and $M_E$.
The matrix $U$ is parameterized as
\begin{align}
	U&=
  \left( 
    \begin{array}{c c c}
      c_{12} c_{13} &
      s_{12} c_{13} &
      s_{13} e^{- i \delta_{\rm CP}} 
      \\
      - c_{23} s_{12} - s_{23} c_{12} s_{13} e^{i \delta_{\rm CP}} &
      c_{23} c_{12} - s_{23} s_{12} s_{13} e^{i \delta_{\rm CP}} &
      s_{23} c_{13} 
      \\
      s_{23} s_{12} - c_{23} c_{12} s_{13} e^{i \delta_{\rm CP}} &
      - s_{23} c_{12} - c_{23} s_{12} s_{13} e^{i \delta_{\rm CP}} &
      c_{23} c_{13}
    \end{array}
  \right)  
  \times
  \mbox{diag} 
  ( 1 \,,~ e^{i \frac{\alpha_{21}}{2} } \,,~ e^{i \frac{\alpha_{31}}{2} }) \,,
\end{align}
where $s_{ij} = \sin \theta_{ij}$ and $c_{ij} = \cos \theta_{ij}$.
$\delta_{\rm CP}$ is the Dirac CP violating phase, 
and $\alpha_{21}$ and $\alpha_{31}$ are the Majorana phases.

It is notable that the model can reproduce the observed values of 
the mixing angles ($\sin^2 \theta_{23}$ is predicted to be larger than 0.54.) 
and the mass squared differences $\Delta m_{ij}^2 = m_i^2 - m_j^2$~\cite{Kobayashi:2018scp}.
Furthermore, the model is very predictive, {\it e.g.},
the normal hierarchy of neutrino masses is predicted,
the Dirac phase is in the range $\delta_{\rm CP} 
=\pm (50^\circ - 180^\circ )$, the effective neutrino mass in the neutrinoless double beta decay is around 22~meV,  and the sum of neutrino masses is larger than 145~meV.
See the details in Ref.~\cite{Kobayashi:2018scp}.

It is natural to verify 
whether the model can explain the BAU or not, 
since it contains all the essential ingredients for the leptogenesis,
{\it i.e.}, right-handed neutrinos, lepton number violation by the 
Majorana masses and CP violation in the modulus field and the coupling constants.
The yield of the BAU depends on the masses $M_i$ and Yukawa coupling constants
of right-handed neutrinos.
Since these parameters are highly restricted due to the symmetry in the model,
we can expect non-trivial relations between the BAU and the observables
in the active neutrino physics, which are the main outcomes of the present article.

%%%%%%%%%%%%%%%%%%%%%%%%%%%%%%%%%%%%%%%%%%%%%%%%%%%%%
%%%%%%%%%%%%%%%%%%%%%%%%%%%%%%%%%%%%%%%%%%%%%%%%%%%%%   
\begin{table}[h]
	\centering
	\renewcommand{\arraystretch}{1.3}
	\begin{tabular}{|c|c|}
	 \hline 
	 	observable & $3\sigma$ range \\ \hline
		$\sin^2 \theta_{12}$ & $0.275 - 0.350$ \\
		$\sin^2 \theta_{23}$ & $0.427 - 0.609$ \\
		$\sin^2 \theta_{13}$ & $0.02046 - 0.02440$ \\
		$\Delta m_{21}^2$ & $(6.79 - 8.01) \times 10^{-5}$~eV$^2$ \\
		$\Delta m_{31}^2$ & $(2.432 - 2.618) \times 10^{-3}$~eV$^2$ \\
		\hline
	\end{tabular}
	\caption{
		The 3$\sigma$ ranges of neutrino oscillation parameters
			for the normal hierarchy case 
 from NuFIT~4.1 (2019)~\cite{Esteban:2018azc}.
		}
	\label{tb:nosc}
\end{table}
%%%%%%%%%%%%%%%%%%%%%%%%%%%%%%%%%%%%%%%%%%%%%%%%%%%%%
%%%%%%%%%%%%%%%%%%%%%%%%%%%%%%%%%%%%%%%%%%%%%%%%%%%%%  
Before discussing the leptogenesis, we shall summarize the properties of right-handed neutrinos inferred from the neutrino oscillation data.
For this purpose, we reanalyze the numerical study of the model 
following Ref.~\cite{Kobayashi:2018scp}.
We use this time the charged lepton  masses in Ref.~\cite{Tanabashi:2018oca}
and update the neutrino oscillation parameters in Ref.~\cite{Esteban:2018azc}
(See Table~\ref{tb:nosc}.).
In addition, we require $ \sum m_i < 160$~meV~\cite{Aghanim:2018eyx}.
We find no qualitative difference from the previous analysis.
Here we show only the results which are essential in the leptogenesis.

\begin{figure}[t]
	\begin{center}
    \includegraphics[height=6cm]{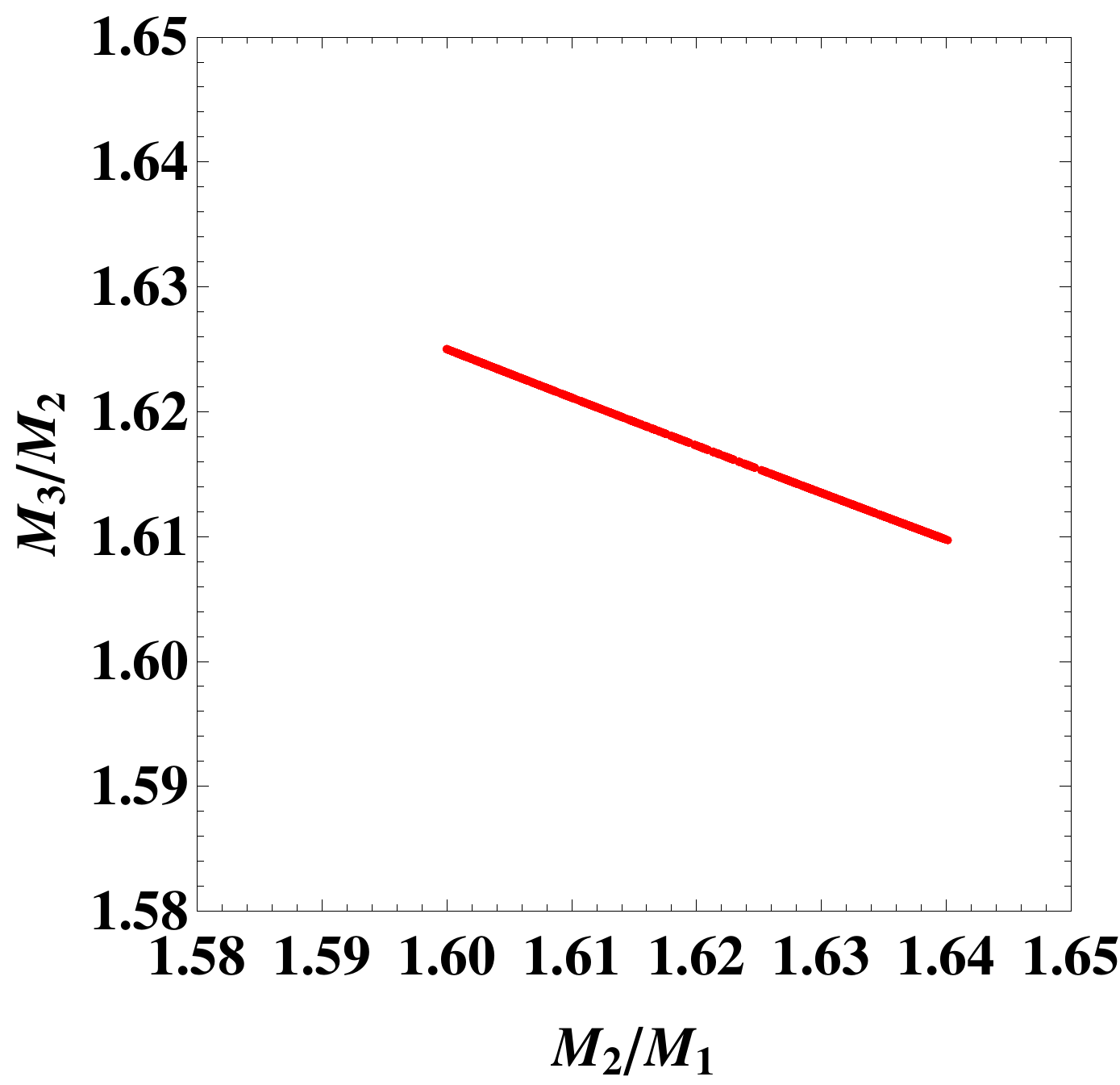}
  \caption{
	The allowed region of mass ratios between right-handed neutrinos.
  }
  \label{RHnumassratio}
  \end{center}
\end{figure}

First, the allowed range of the mass ratios of right-handed neutrinos
is shown in Fig.~\ref{RHnumassratio}.
It is seen that $M_2/M_1$ and $M_3/M_2$ are both about 1.6.   
Notice that the absolute values of right-handed neutrino masses
cannot be determined from the oscillation data, however, 
as we will show in section~\ref{sec:yieldBAU} the order of magnitude of them can be found from the BAU.
There are two consequences to the leptogenesis;
(1) All three right-handed neutrinos should be taken into account 
in the leptogenesis dynamics.
(2) The resonant production of the lepton asymmetry by the decays \cite{Pilaftsis:2003gt}
is less effective.
%%%%%%%%%%%%%%%%%%%%%%%%%%%%%%%%%%%%%%%%%%%%%%%%%%%%%%%%%%%%%%%%%%%%%% 
%%%%% ** Figure ** %%%%%%%%%%%%%%%%%%%%%%%%%%%%%%%%%%%%%%%%%%%%%%%%%%%
%%%%%%%%%%%%%%%%%%%%%%%%%%%%%%%%%%%%%%%%%%%%%%%%%%%%%%%%%%%%%%%%%%%%%%

%%%%%%%%%%%%%%%%%%%%%%%%%%%%%%%%%%%%%%%%%%%%%%%%%%%%%%%%%%%%%%%%%%%%%% 
%%%%% ** Figure ** %%%%%%%%%%%%%%%%%%%%%%%%%%%%%%%%%%%%%%%%%%%%%%%%%%%
\begin{figure}[h]
	\begin{center}
    \includegraphics[height=6cm]{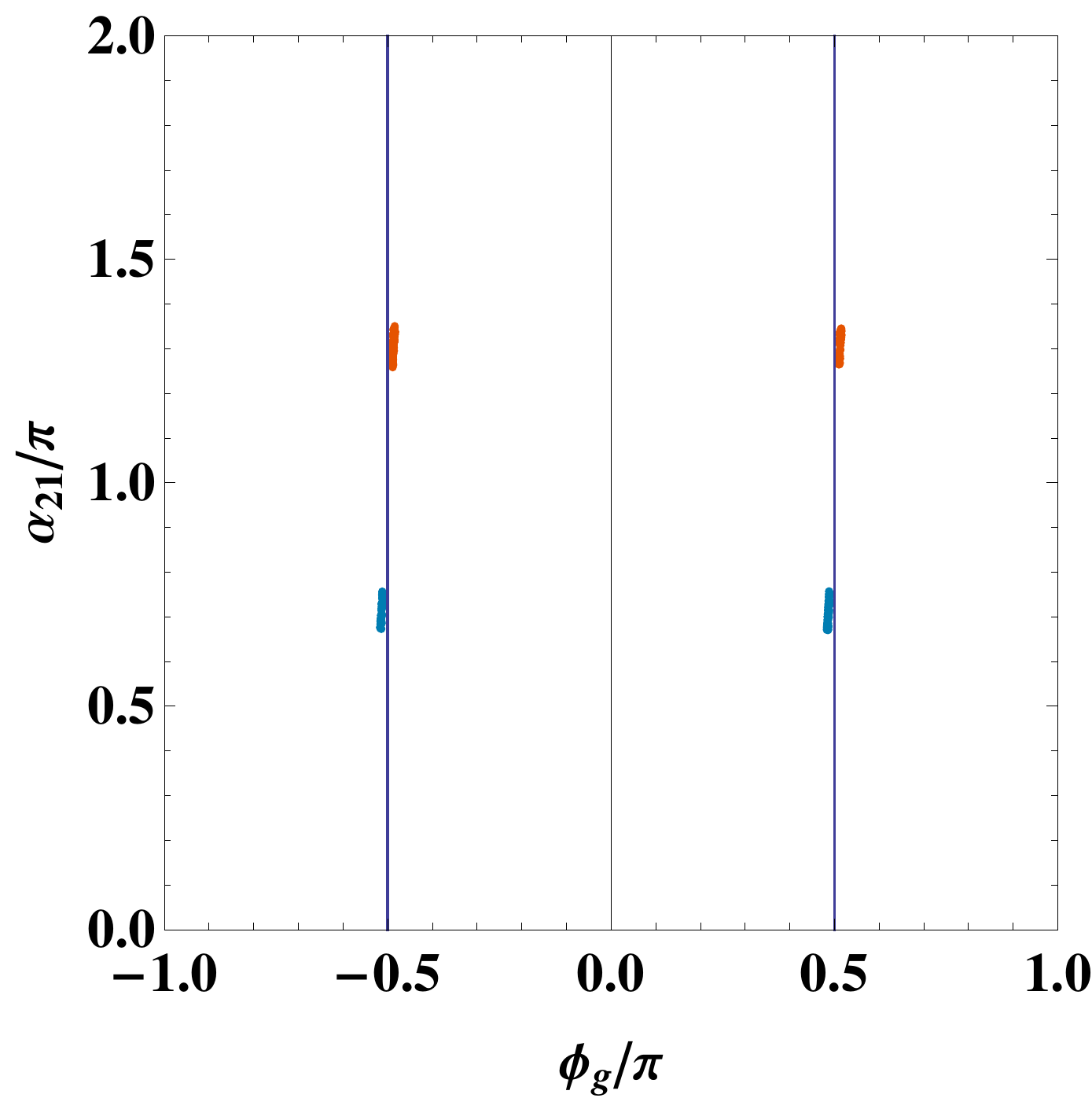}
    \hspace{1cm}
    \includegraphics[height=6cm]{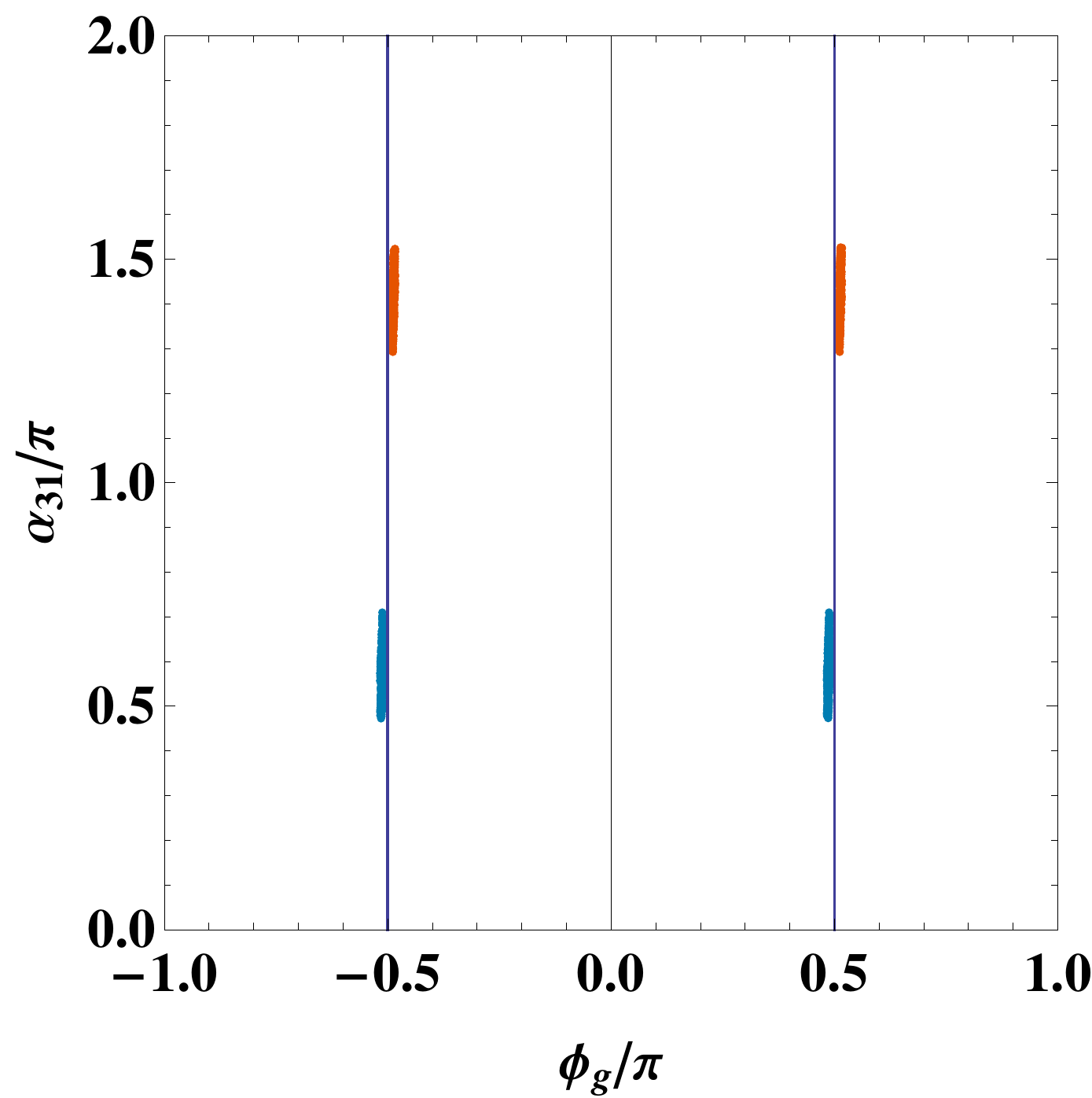}
  \caption{
  	The allowed regions of $\alpha_{21}$ and $\phi_g$ (left)
  	and that of $\alpha_{31}$ and $\phi_g$ (right).
  	At the orange (blue) points the sign of $Y_B$ is positive (negative).
  }
  \label{CPV_Majorana}
  \end{center}
	\begin{center}
    \includegraphics[height=6cm]{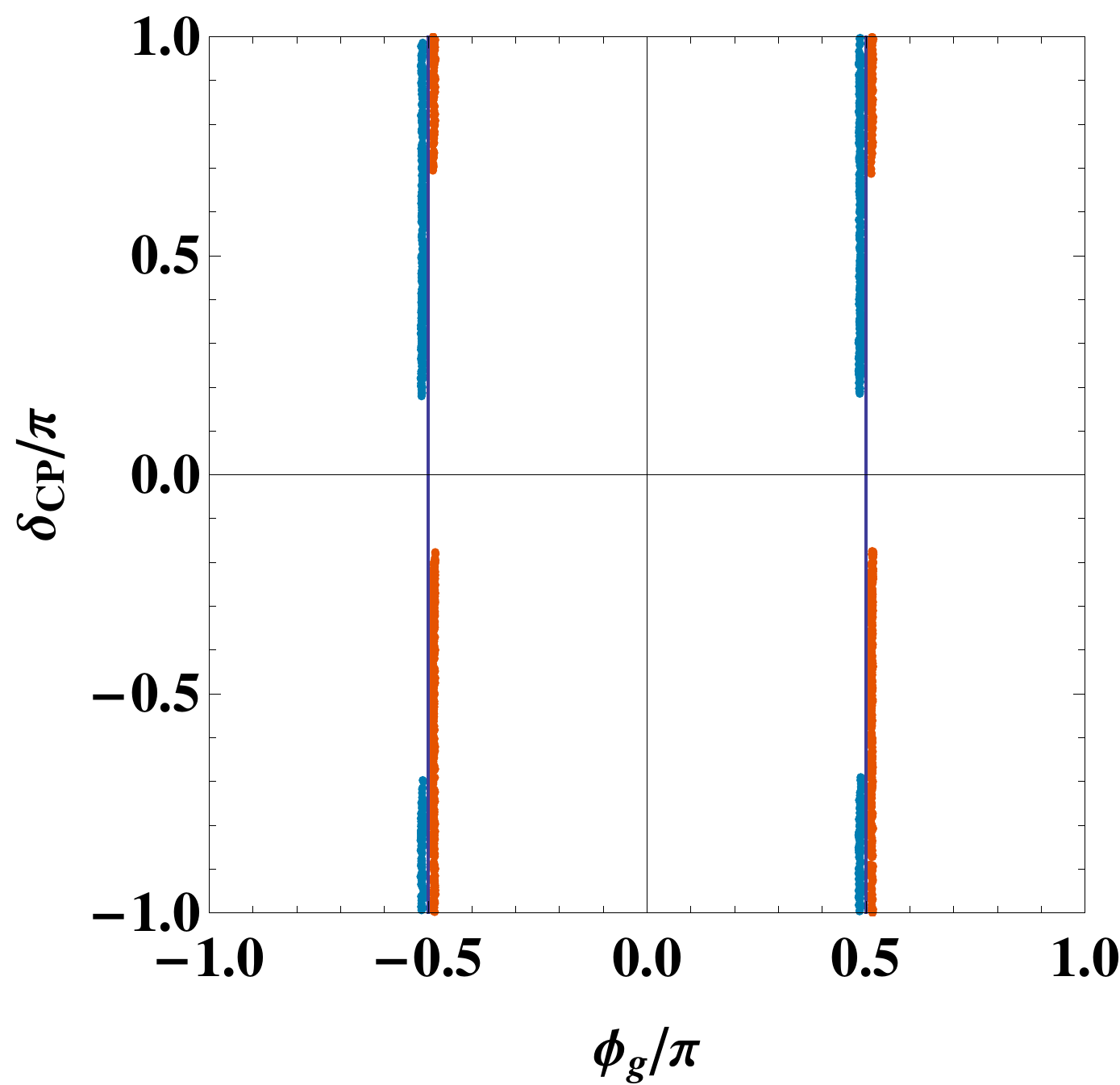}
  \caption{
	  The allowed regions of $\delta_{\rm CP}$ and $\phi_g$.
  	At the orange (blue) points the sign of $Y_B$ is positive (negative).
  }
  \label{CPV_delta}
  \end{center}
\end{figure}
%%%%%%%%%%%%%%%%%%%%%%%%%%%%%%%%%%%%%%%%%%%%%%%%%%%%%%%%%%%%%%%%%%%%%%

Next, we consider the CP violating parameters in the present model.
There are two such parameters $\tau$ and $g_2$, which are relevant for baryogensis.   
Note that they induce the CP violations for active neutrinos (that
are observables at low energies) and also for right-handed neutrinos.
Interestingly, we observe the strong correlations between 
the Majorana phases $\alpha_{21, 31}$ and the phase of $g_2$, $\phi_g$,
which is represented in Fig.~\ref{CPV_Majorana}.
The allowed value of the Majorana phase changes whether 
$\phi_g$ is slightly larger or smaller than $\pm \pi/2$.
No strong correlation is found between the Majorana phases and $\tau$.
On the other hand, the Dirac phase $\delta_{\rm CP}$ depends non-trivially 
on $\tau$ and $\phi_g$, and there is no correlation between these parameters.

%%%%%%%%%%%%%%%%%%%%%%%%%%%%%%%%%%%%%%%%%%%%%%%%%%%%%%%%%%%%%%%%%%%%%%
%%%%%%%%%%%%%%%%%%%%%%%%%%%%%%%%%%%%%%%%%%%%%%%%%%%%%%%%%%%%%%%%%%%%%%
%%%%%%%%%%%%%%%%%%%%%%%%%%%%%%%%%%%%%%%%%%%%%%%%%%%%%%%%%%%%%%%%%%%%%%
\section{Leptogenesis}
Now we are at the point to discuss the leptogenesis by decays of right-handed neutrinos
in the model.  
As explained before, the mass ratios of right-handed neutrinos are not so large
as $\sim 1.6$,  and then 
we have to include the effects of all three right-handed neutrinos to the leptogenesis.
Here we assume for simplicity that the reheating temperature of inflation
is sufficiently higher than the mass of the heaviest right-handed neutrino
and that the initial abundances of all right-handed neutrinos are zero.
On the other hand, the mass differences of right-handed neutrinos are not so small,
and then the resonant enhancement of the leptogenesis \cite{Pilaftsis:2003gt} does not occur.
Thus, we shall use the formalism based on the Boltzmann equations for the estimation of the asymmetries.
Furthermore, as shown below, the required masses of right-handed 
neutrinos are ${\cal O}(10^{13})$~GeV, and hence the simple one-flavor approximation
of the leptogenesis can be applied and we only consider the (total) 
lepton asymmetry neglecting the so-called flavor effect~\cite{Abada:2006fw,Nardi:2006fx,Abada:2006ea,Blanchet:2006be,Pascoli:2006ie,Pascoli:2006ci,Moffat:2018smo,DeSimone:2006nrs}.

We solve the Boltzmann equations for the number densities
$n_{N_I}$ for right-handed neutrinos and the lepton asymmetry density $n_L$.
\begin{align}
	\frac{d Y_{N_I}}{dz}	
		&=	
		\frac{- z}{s H(M_1)}	\Bigg\{	
		\left(	\frac{Y_{N_I}}{Y_{N_I}^{eq}}	-	1	\right) 	
		\left(\gamma_{N_I}+2 \gamma_{tI}^{(3)}	+4\gamma_{tI}^{(4)} \right)	
		+
		\sum_{J=1}^3	
			\left( \frac{Y_{N_I}}{Y_{N_I}^{eq}}\frac{Y_{N_J}}{Y_{N_J}^{eq}}	-	1	\right)		
			\left( \gamma_{N_I N_J}^{(2)}+\gamma_{N_I N_J}^{(3)} \right)
			\Bigg\}	\,,
			\label{eq:yieldNI}
	\\
	\frac{d Y_{L}}{dz}	
	&=	
	\frac{-z}{s H(M_1)}	\Bigg\{
	\sum_{I=1}^3 \left[	
		\left( 1 - \frac{Y_{N_I}}{Y_{N_I}^{eq}} \right) \varepsilon_I \, \gamma_{N_I} 
		+ \frac{Y_{L}}{Y_{\ell}^{eq}} \frac{\gamma_{N_I}}{2}
	\right]	
	+	
	\frac{Y_{L}}{Y_{\ell}^{eq}}	\left( 2 \gamma_{N}^{(2)}	+ 2\gamma_{N}^{(13)} \right)
	\nonumber \\
	&~~~~~~~~~+	
	\frac{Y_{L}}{Y_{\ell}^{eq}}	
	\sum_{I=1}^3
	\left[ \frac{Y_{N_I}}{Y_{N_I}^{eq}}\gamma_{tI}^{(3)} 
		+ 2\gamma_ {tI}^{(4)}
		+ \frac{Y_{N_I}}{Y_{N_I}^{eq}} 
		\left( \gamma_{WI}^{(1)} +\gamma_{BI}^{(1)} \right)
		+
    \gamma_{WI}^{(2)} +\gamma_{WI}^{(3)}+ \gamma_{BI}^{(2)} +\gamma_{BI}^{(3)} 
  \right]
  \Bigg\}	\,,
  \label{eq:yieldYL2}
\end{align}
where $z =M_1/T$.  
The yields are defined by 
$Y_{N_I} = n_{N_I}/s$ and $Y_{L} = n_L /s$ 
with the entropy density of the universe $s$. 
The superscript "$eq$" represents its equilibrium value. We take the yield for a massless particle with one degree of freedom in equilibrium as $Y_{\ell}^{eq}$. Here we apply the Boltzmann approximation and $Y_{\ell}^{eq} = 45/(2\pi^4 g_{\ast s})$, with $g_{\ast s}=110.75$.
Our notations of the reaction densities correspond to those in 
Ref.~\cite{Plumacher:1998ex}. The $\mathrm {CP}$ asymmetry parameter for the leptogenesis $\varepsilon_I$ is defined by
\begin{align}
\varepsilon_{I}
=\frac{\Gamma\left(N_{I} \rightarrow L+\bar H_u\right)-\Gamma\left(N_{I} \rightarrow \bar L +H_u\right)}{\Gamma\left(N_{I} \rightarrow L+\bar H_u\right)+ \Gamma\left(N_{I} \rightarrow \bar L+H_u\right).
}
\end{align}
The explicit form of the reaction density 
for the $N_I$ decay is given by
\begin{align}
	&\gamma_{N_I} 
	= \frac{\left(Y_\nu {Y_\nu}^\dagger \right)_{I I}}{8 \pi^{3}} 
	M_{1}^{4} a_{I}^{3/2}
	\frac{K_{1}\left(\sqrt{a_{I}} z\right)}{z} \,,
\end{align}
where $z =M_1/T$, $a_I = (M_I / M_1)^2$, and $K_1(x)$ is the modified Bessel function
of the second kind.
Note that $Y_\nu$ is the Yukawa coupling matrix of neutrinos in the base where
the mass matrices of charged leptons and right-handed neutrinos are diagonal.
The reaction density for the process $A + B \to C + D$ is given by
\begin{align}
	&\gamma(A+B \rightarrow C+D) 
	= \frac{T}{64 \pi^{4}}
	\int_{ \left(m_{A}+m_{B}\right)^{2} }^{\infty} 
	ds \, 
	\hat{\sigma}(s) \sqrt{s} K_{1}\left(\frac{\sqrt{s}}{T}\right)\,,
\end{align} 
where $m_{A}$ and $m_B$ are masses of the initial particles 
and $\hat \sigma (s)$ denotes the reduced cross section for the process.
As for the $\Delta L=1$ processes induced through top Yukawa intraction,
the $\Delta L=2$ scattering processes and 
the annihilation processes of right-handed neutrinos, the expressions of the reduced cross sections are found in Ref.~\cite{Plumacher:1998ex}. 
Note that the correct subtraction of the $N_I$ on-shell contribution 
for $L H_u \to \bar L \bar H_u$~process gives~\cite{Giudice:2003jh}
\begin{align}
  \hat{\sigma}_{N}^{(2)}(x)	
  &=	
  \frac{1}{2\pi} 
  \Bigg[ \sum_{I} (Y_\nu {Y_\nu}^\dagger)_{II}^2	\frac{a_I}{x} 
  \Big\{\frac{x}{a_I}+\frac{x}{D_{I}}	
  - \Big(1+\frac{x+a_I}{D_{I}} \Big)
  \log \Big(	\frac{x+a_I}{a_I}\Big)	
  \Big\}	
  \nonumber \\
	&~~~
	+\sum_{I>J}	\mathrm{Re}[(Y_\nu {Y_\nu}^\dagger)_{IJ}^2]
	\frac{\sqrt{a_I a_J}}{x} 
	\Big\{ 	
  \frac{x^2 + x ( D_I + D_J )}{D_{I} D_{J}} 
	+ 
	(x+a_I) \Big(\frac{2}{a_J-a_I}-\frac{1}{D_J}\Big) 
	\ln\Big(\frac{x+a_I}{a_I}\Big)
	\nonumber 	\\
	&\hspace{5cm}
	+
	(x+a_J)
	\Big(\frac{2}{a_I-a_J}-\frac{1}{D_I} \Big) 
	\ln \Big( \frac{x+a_J}{a_J} \Big)	\Big\} \Bigg]	\,,
\end{align}
where $c_I = (\Gamma_{N_I}/M_1)^2$
and $D_I = [(x-a_I)^2 + a_I c_I]/(x-a_I)$, in which $\Gamma_{N_I}$ is the total decay rate of right-handed neutrino $N_I$.
The reduced cross sections for $\Delta L=1$ processes through 
the $SU(2)_L$ SM gauge interaction are~\cite{Giudice:2003jh,Pilaftsis:2003gt}
\begin{align}
	\hat \sigma_{WI}^{(1)}(x)	
	&=	
	\frac{3g_2^2(Y_\nu {Y_\nu}^\dagger)_{II}}{16\pi x^2}
  \Big[ -2x^2+6a_I x -4a_I^2	+(x^2-2a_I x +2a_I^2) 
  \ln \Big|\frac{x-a_I+a_L}{a_L} \Big|
  \nonumber \\
	&\hspace{3cm}	
	+
		\frac{x(a_L x +a_L a_I -a_W a_I)(a_I -x)}{a_L(x-a_I+a_L)}
	\Big] \,,
	\\
	\hat \sigma_{WI}^{(2)}(x)	
	&=
	\frac{3g_2^2(Y_\nu {Y_\nu}^\dagger)_{II}}{8\pi x(x-a_I)}
  \Big[
  	2a_I x \ln \Big| \frac{x -a_I +a_H}{a_H} \Big|	
	+ (x^2 +a_I^2) \ln \Big| \frac{x-a_I-a_W-a_H}{-a_W -a_H} \Big|
	\Big] \,,
	\\
	\hat \sigma_{WI}^{(3)}(x)	
	&=	
	\frac{3g_2^2(Y_\nu {Y_\nu}^\dagger)_{II} a_I}{16\pi x^2}
	\Big[	\frac{x^2 -4a_I x +3a_I^2}{a_I} 
		+ 4(x-a_I) \ln \Big| \frac{x-a_I +a_H}{a_H} \Big|	
	  - \frac{x(4a_H -a_W)(x-a_I)}{a_H(x-a_I+m_H)}	
	  \Big] \,.
\end{align}
Here $\hat \sigma_{WI}^{(1)}$, $\hat \sigma_{WI}^{(2)}$ and $\hat \sigma_{WI}^{(3)}$ are the reaction densities for the processes $N_I L \rightarrow H_u W$, $N_I W \rightarrow \bar L H_u$ and $N_I \bar H_u \rightarrow \bar L W$, respectively.
We have used $a_{L, H, W, B} = m_{L,H_u, W, B}^2/M_1^2$
where
$m_X$ with $X = L, H_u, W, B$ are thermal masses of lepton doublets, up-type Higgs, $SU(2)_L$ gauge bosons and $U(1)_Y$ gauge boson, respectively.
The reaction densities for the $\Delta L = 1$ processes
through $U(1)_Y$ gauge interaction are obtained by substituting 
$a_W \to a_B$ and $\frac{3}{2} g_2^2 \to \frac{1}{4} g_Y^2$
in $\hat \sigma_{WI}^{(i)}$.  

For the estimation of the reaction densities, we have taken into account
the one-loop RGE evolutions of couplings and the renormalization scale
is taken as $\mu = 2\pi T$.  The important effect is the suppression of top Yukawa coupling
at high temperatures due to the RGE effect, which reduces the washout of the produced lepton asymmetry and enlarges a viable parameter space.

The Boltzmann equations are then solved numerically and the total lepton asymmetry 
$Y_L$ from the decays of right-handed neutrinos is estimated.  
The present baryon asymmetry can be estimated 
as $Y_B= -8/23 Y_L$,
\footnote{Here, we assume that two-Higgs doublet survive at sphaleron freeze-out temperature. On the other hand, $Y_B = -28/79 Y_L$ for the one-Higgs doublet case.}  where we have taken into account for 
the effect of the two Higgs doublets.

Now, since the lightest right-handed neutrino is sufficiently heavy, we can neglect 
the flavor effect of the leptogenesis~\cite{Abada:2006fw,Nardi:2006fx,Abada:2006ea,Blanchet:2006be,Pascoli:2006ie,Pascoli:2006ci,Moffat:2018smo,DeSimone:2006nrs}.  In this case, the final baryon asymmetry becomes
insensitive to the PMNS mixing matrix of active neutrinos.   However, in the considering 
model, the phases in the PMNS matrix and the high energy phases associated with 
right-handed neutrinos are originated in the limited complex parameters 
$\tau$ and $g_2$.  In this situation, there may exist the correlations between 
the phases in the PMNS matrix and the yield of the BAU.

%%%%%%%%%%%%%%%%%%%%%%%%%%%%%%%%%%%%%%%%%%%%%%%%%%%%%%%%%%%%%%%%%%%%%%
%%%%%%%%%%%%%%%%%%%%%%%%%%%%%%%%%%%%%%%%%%%%%%%%%%%%%%%%%%%%%%%%%%%%%%
%%%%%%%%%%%%%%%%%%%%%%%%%%%%%%%%%%%%%%%%%%%%%%%%%%%%%%%%%%%%%%%%%%%%%%
\section{Sign and magnitude of baryon asymmetry}\label{sec:yieldBAU}
Let us then show the results of the BAU by right-handed neutrinos in the model.
We begin with the sign of the BAU produced by right-handed neutrinos
in the model.   

\begin{figure}[t]
	\begin{center}
    \includegraphics[height=6cm]{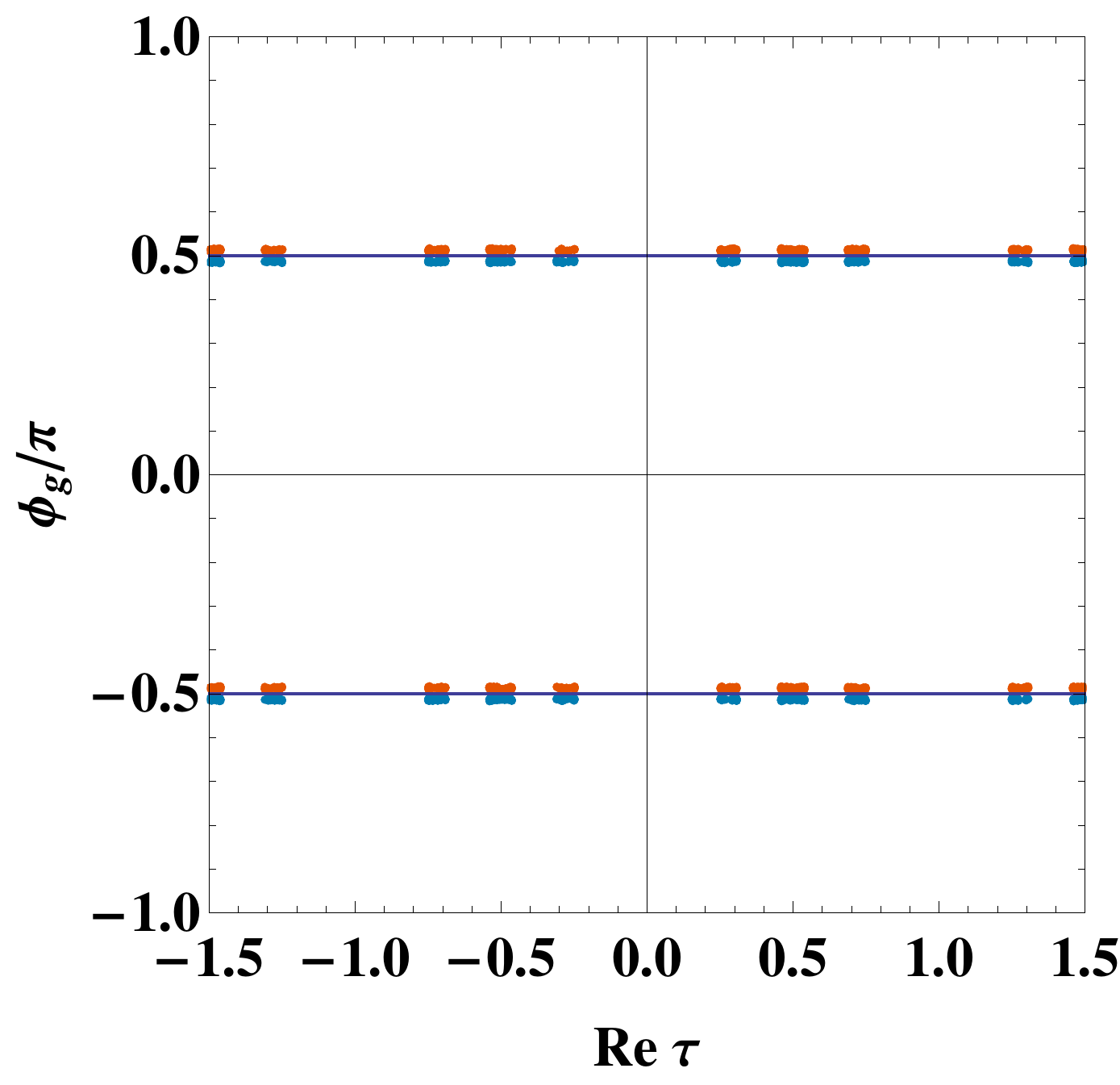}
    \caption{
    	The allowed regions of Re $\tau$ and phase $\phi_g$.	  
    Red and blue points correspond to the positive and negative signs of the BAU, respectively.
  }
  \label{RetauPhig}
  \end{center}
	\begin{center}
    \includegraphics[height=6cm]{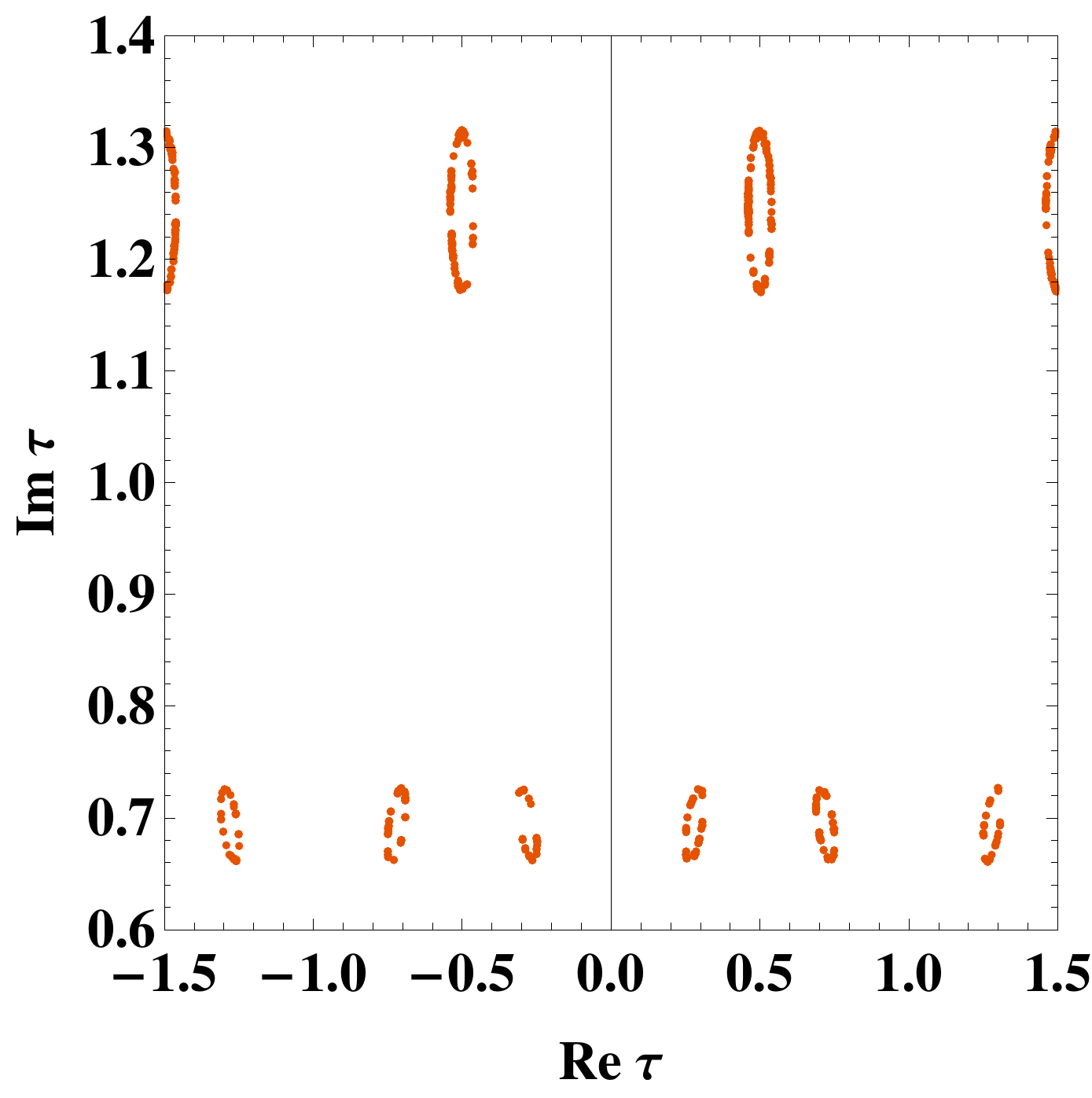}
    \hspace{1cm}
    \includegraphics[height=6cm]{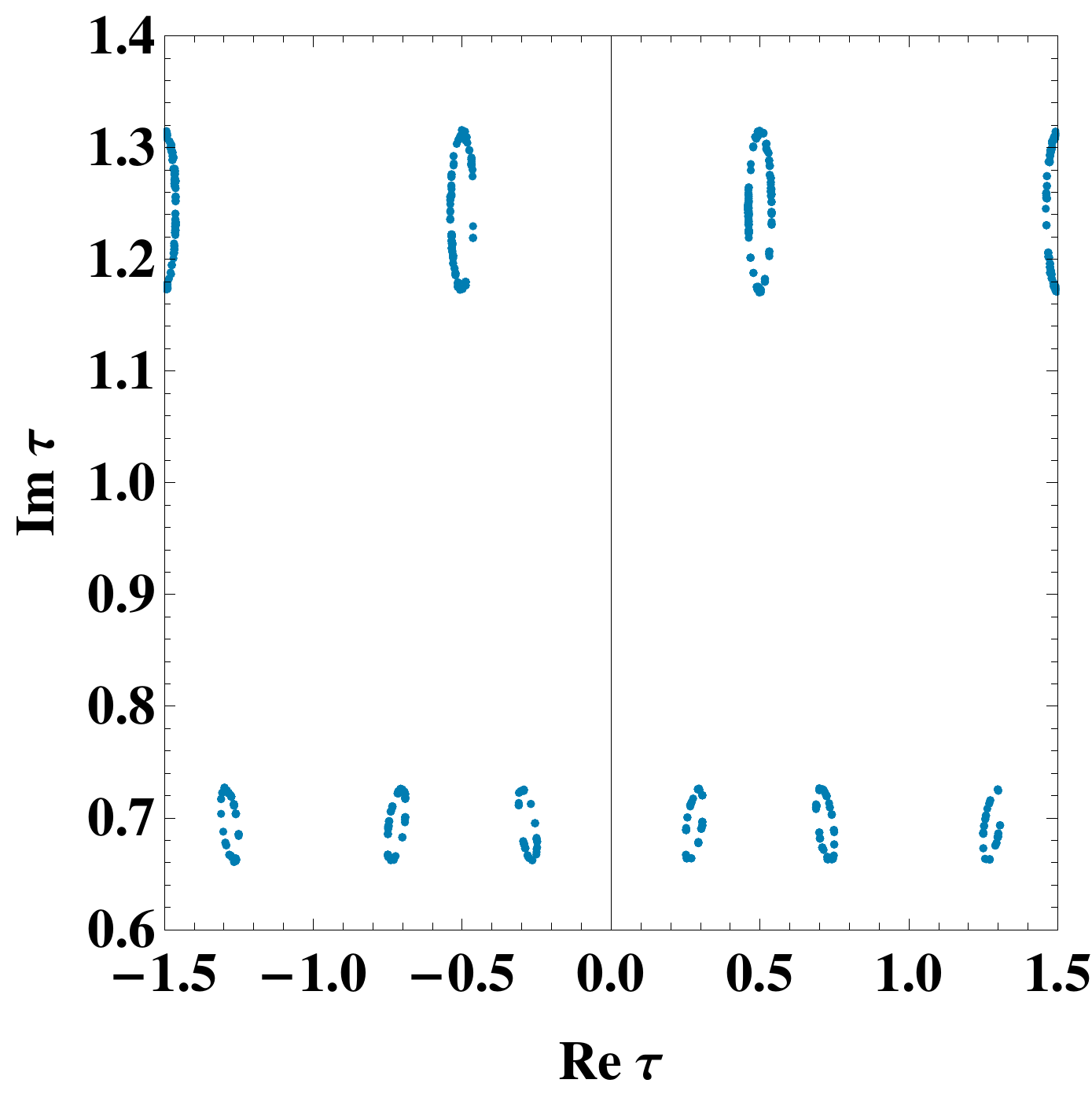}
    \caption{
    	The allowed regions of Re $\tau$ and Im $\tau$
		when the sign of the BAU is positive (left) or negative (right).
  }
  \label{RetauImtau}
  \end{center}
\end{figure}

The first important result is that 
the sign of the BAU is determined by the phase $\phi_g$ of the complex coupling $g_2$.
This point is represented in Fig.~\ref{RetauPhig}.
The positive BAU is obtained when $\phi_g$ is slightly larger than
$\pi/2$ or $- \pi/2$.  
On the other hand,  as shown in Fig.~\ref{RetauImtau},
there is no strong correlation between the BAU sign and 
the complex parameter $\tau$.
\footnote{The mass ratios between right-handed neutrinos are determined by $\tau$ as shown in Eq.~(\ref{MajoranaR}). It is then found from Fig.~\ref{RetauImtau} there is no correlation between mass hierarchy of right-handed neutrinos and the sign of the BAU.}
%%%%%%%%%%%%%%%%%%%%%%%%%%%%%%%%%%%%%%%%%%%%%%%%%%%%%%%%%%%%%%%%%%%%%% 
%%%%% ** Figure ** %%%%%%%%%%%%%%%%%%%%%%%%%%%%%%%%%%%%%%%%%%%%%%%%%%%
%%%%%%%%%%%%%%%%%%%%%%%%%%%%%%%%%%%%%%%%%%%%%%%%%%%%%%%%%%%%%%%%%%%%%%

\begin{figure}[t]
	\begin{center}
    \includegraphics[height=6cm]{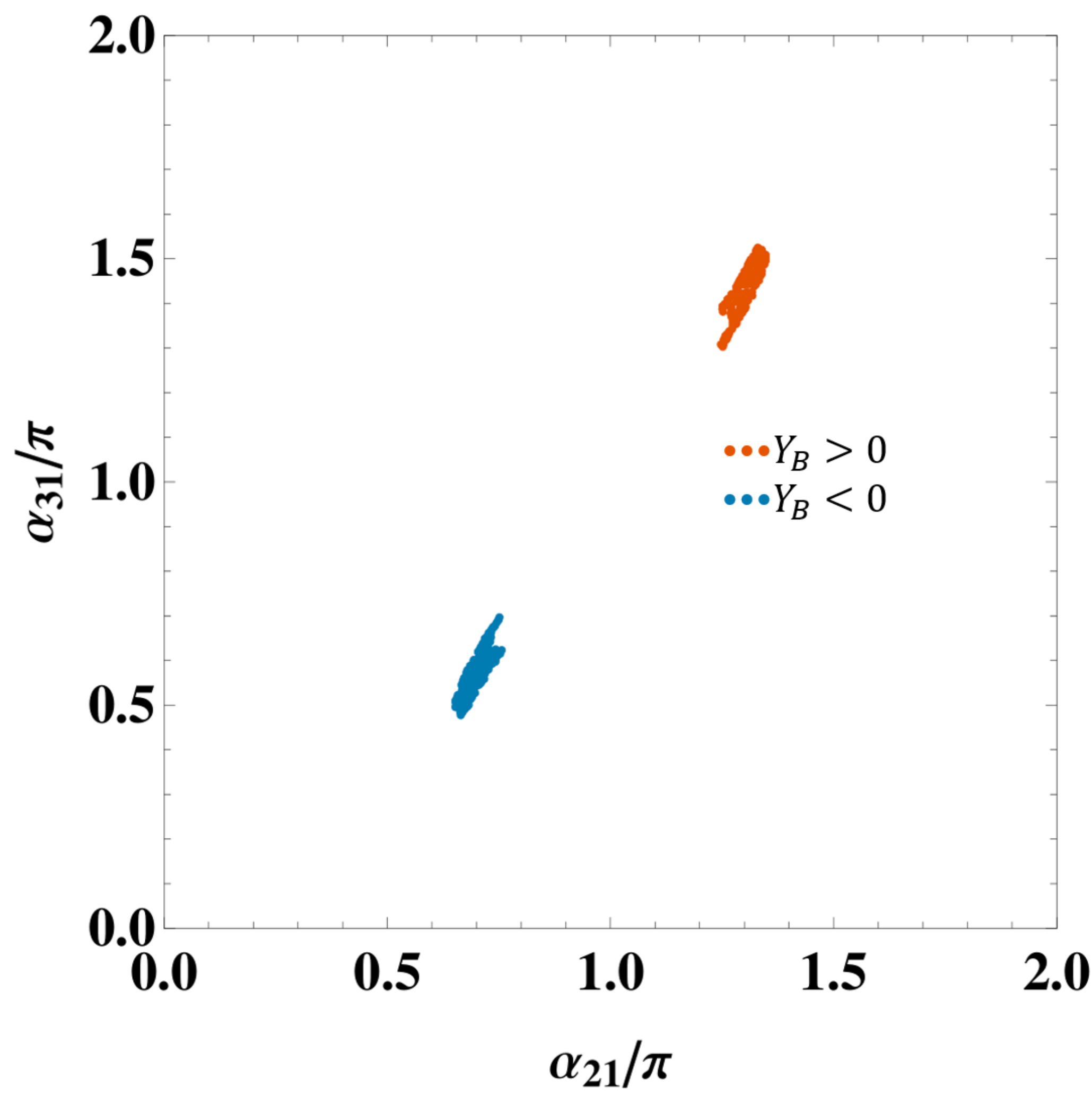}
    \hspace{1cm}
    \includegraphics[height=6cm]{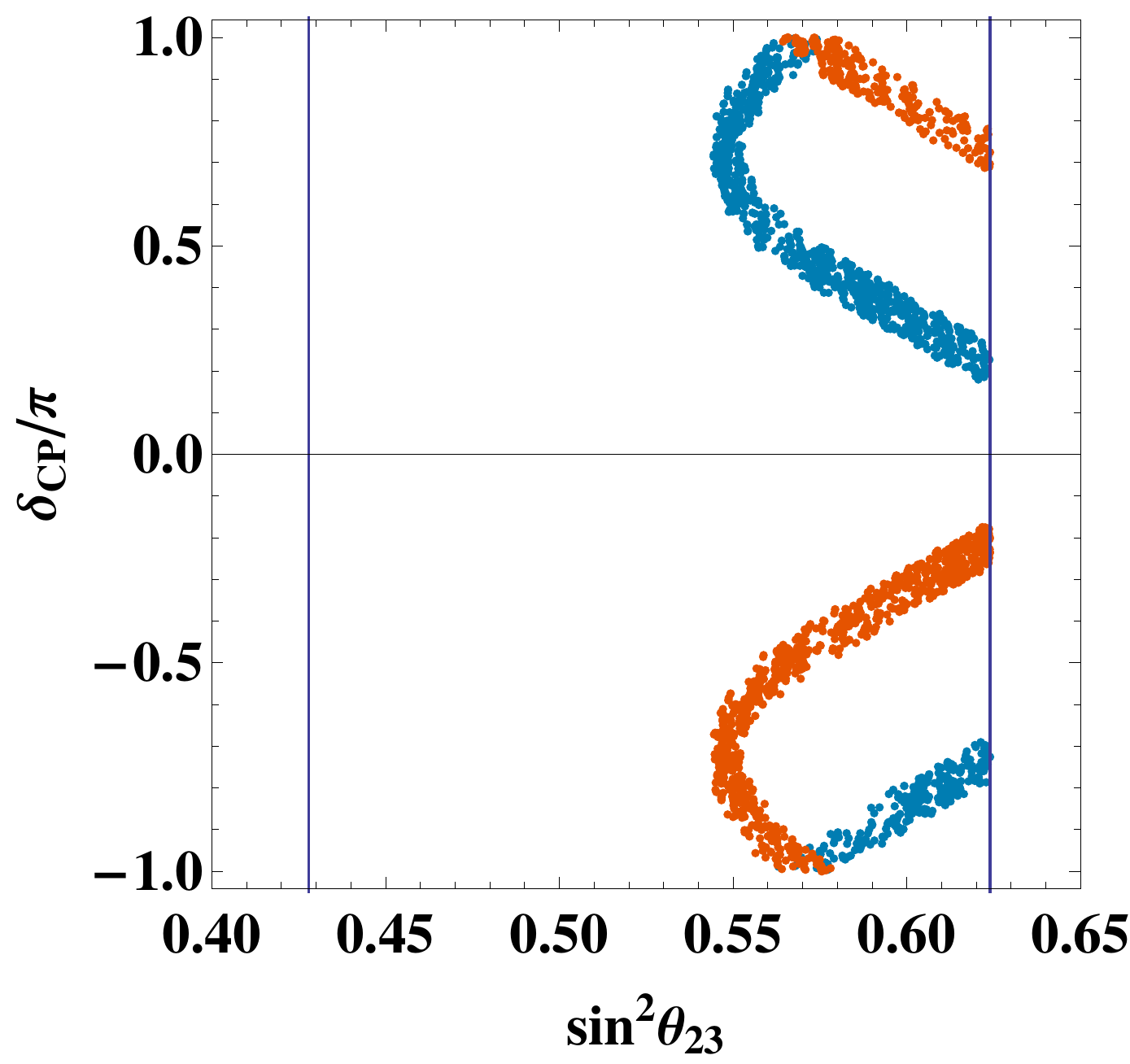}
  \caption{
		 $\alpha_{21}$ and $\alpha_{31}$ (left),
		 and $\sin^2 \theta_{23}$ and $\delta_{\rm CP}$ (right).
		$Y_B$ is positive (negative) at the orange (blue) points.
  }
  \label{AlpAlp}
  \end{center}
\end{figure}

\begin{figure}[t]
	\begin{center}
    \includegraphics[height=6cm]{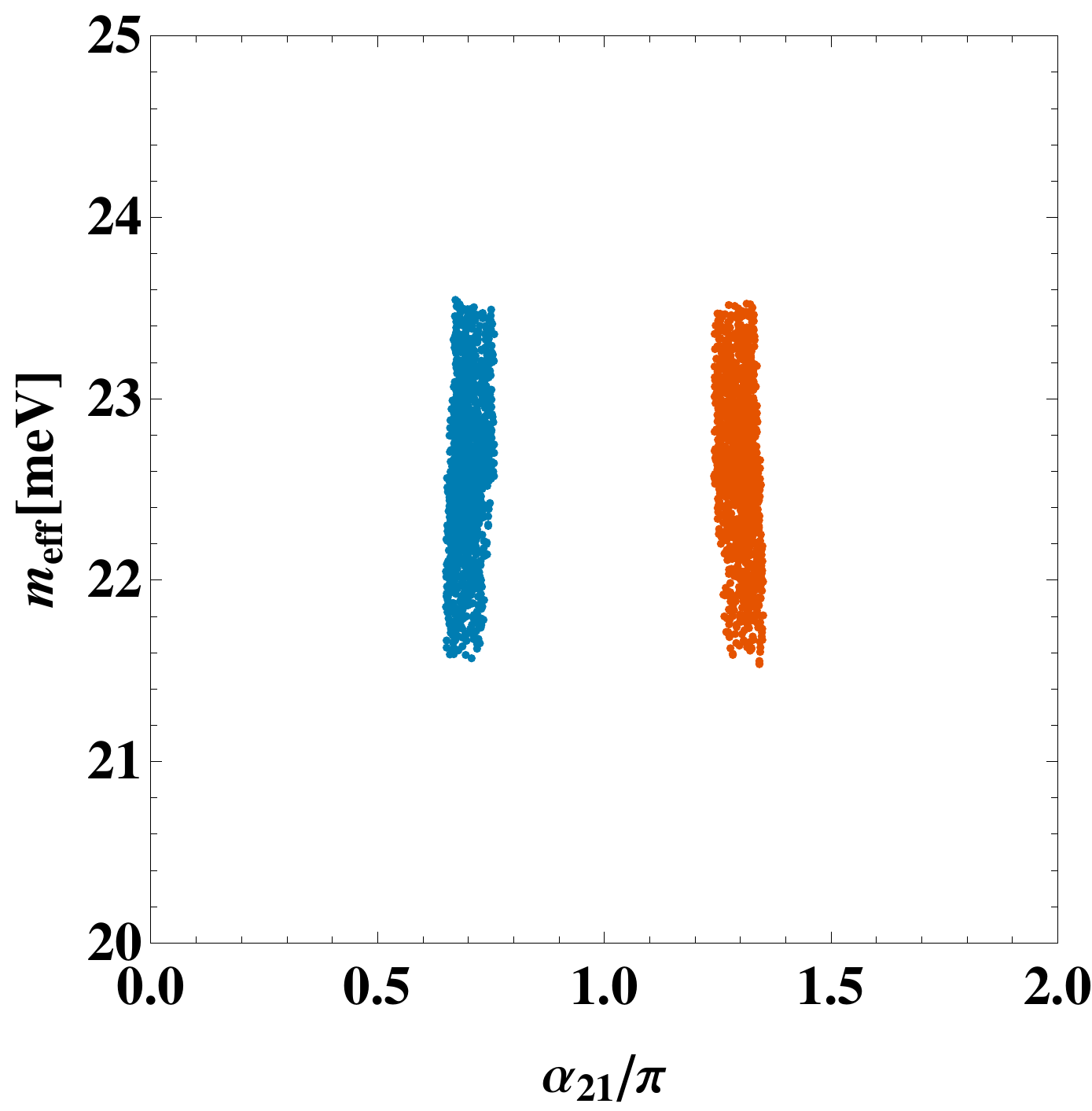}
    \caption{
The allowed ranges of $\alpha_{21}$ and $m_{\rm eff}$.
  }
  \label{alp31vsmeff}
  \end{center}
\end{figure}

The phase $\phi_g$ is strongly correlated with Majorana phases $\alpha_{21}$ and 
$\alpha_{31}$, and then the positive BAU is possible only for 
the restricted range of Majorana phases, which is shown in Fig.~\ref{AlpAlp}.
We can see that two regions in the Majorana phases
are allowed by the observational data about active neutrinos,
which are related as $\alpha_{21, 31} \simeq 2 \pi - \alpha_{21,31}$.
The positive BAU is, however,  realized only when 
$\alpha_{21} \sim 1.3~\pi$ and $\alpha_{31} \sim 1.5~\pi$.
This is an important prediction of the leptogenesis in the model although 
the experimental measurements of Majorana phases are very difficult.
Notice that the range of the effective neutrino mass
in the $0 \nu \beta \beta$ decay is the same for the both cases
$Y_B >0$ and $Y_B <0$, as shown in Fig.~\ref{alp31vsmeff}.

%%%%%%%%%%%%%%%%%%%%%%%%%%%%%%%%%%%%%%%%%%%%%%%%%%%%%%%%%%%%%%%%%%%%%% 
%%%%% ** Figure ** %%%%%%%%%%%%%%%%%%%%%%%%%%%%%%%%%%%%%%%%%%%%%%%%%%%
%%%%%%%%%%%%%%%%%%%%%%%%%%%%%%%%%%%%%%%%%%%%%%%%%%%%%%%%%%%%%%%%%%%%%%
%%%%%%%%%%%%%%%%%%%%%%%%%%%%%%%%%%%%%%%%%%%%%%%%%%%%%%%%%%%%%%%%%%%%%%

Second, we find from Fig.~\ref{AlpAlp} that
the dependence of the BAU sign on Dirac phase is different 
depending on $\sin^2 \theta_{23}$.
For $\sin^2 \theta_{23} \lesssim 0.58$
the positive BAU can be obtained for $\delta_{\rm CP} < 0$.
On the other hand, for $\sin^2 \theta_{23} \gtrsim 0.58$
the positive BAU is possible for both $\delta_{\rm CP} <0$ and $\delta_{\rm CP}>0$.
This shows that the precise measurements of $\sin^2 \theta_{23}$
and $\delta_{\rm CP}$ provide a crucial test for the correct sign of the BAU 
in the considering baryogenesis scenario.

\begin{figure}[t!]
	\begin{center}
    \includegraphics[height=6cm]{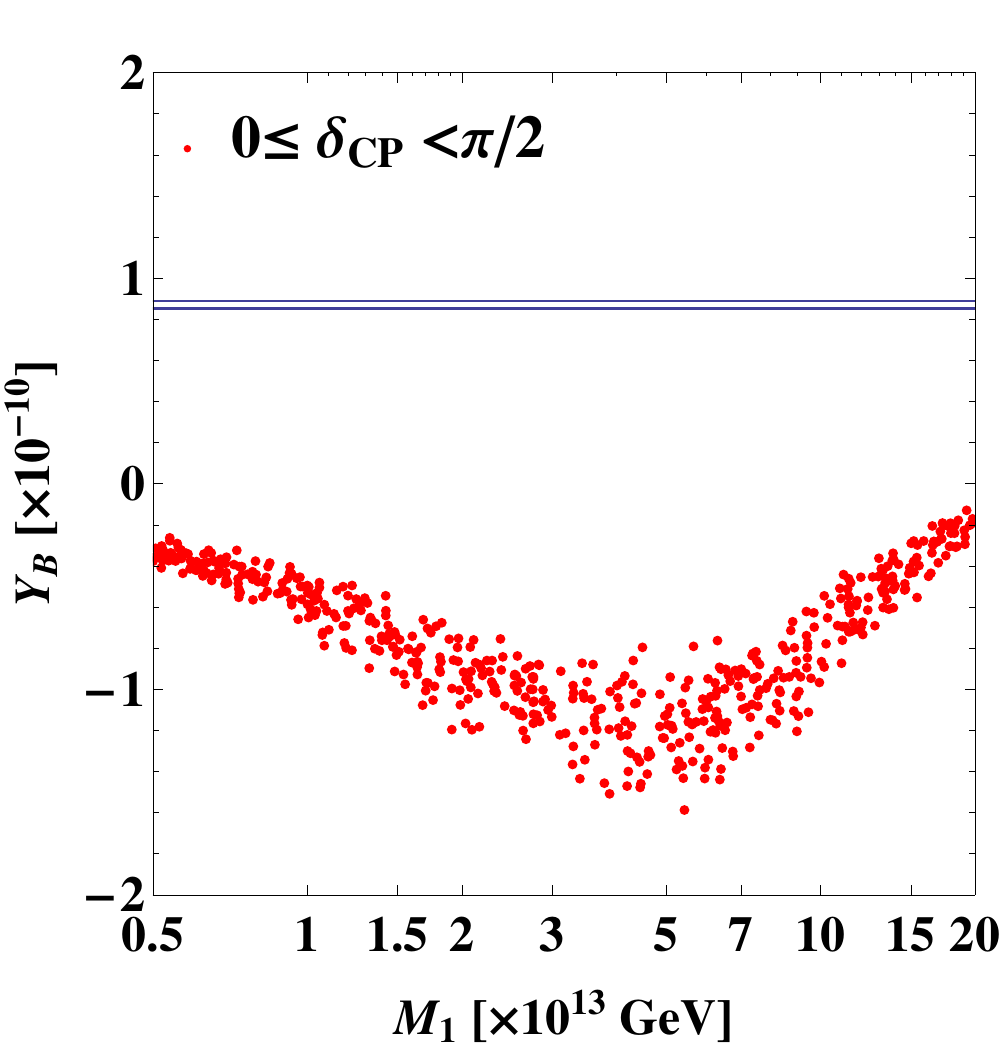}
    \hspace{1cm}
    \includegraphics[height=6cm]{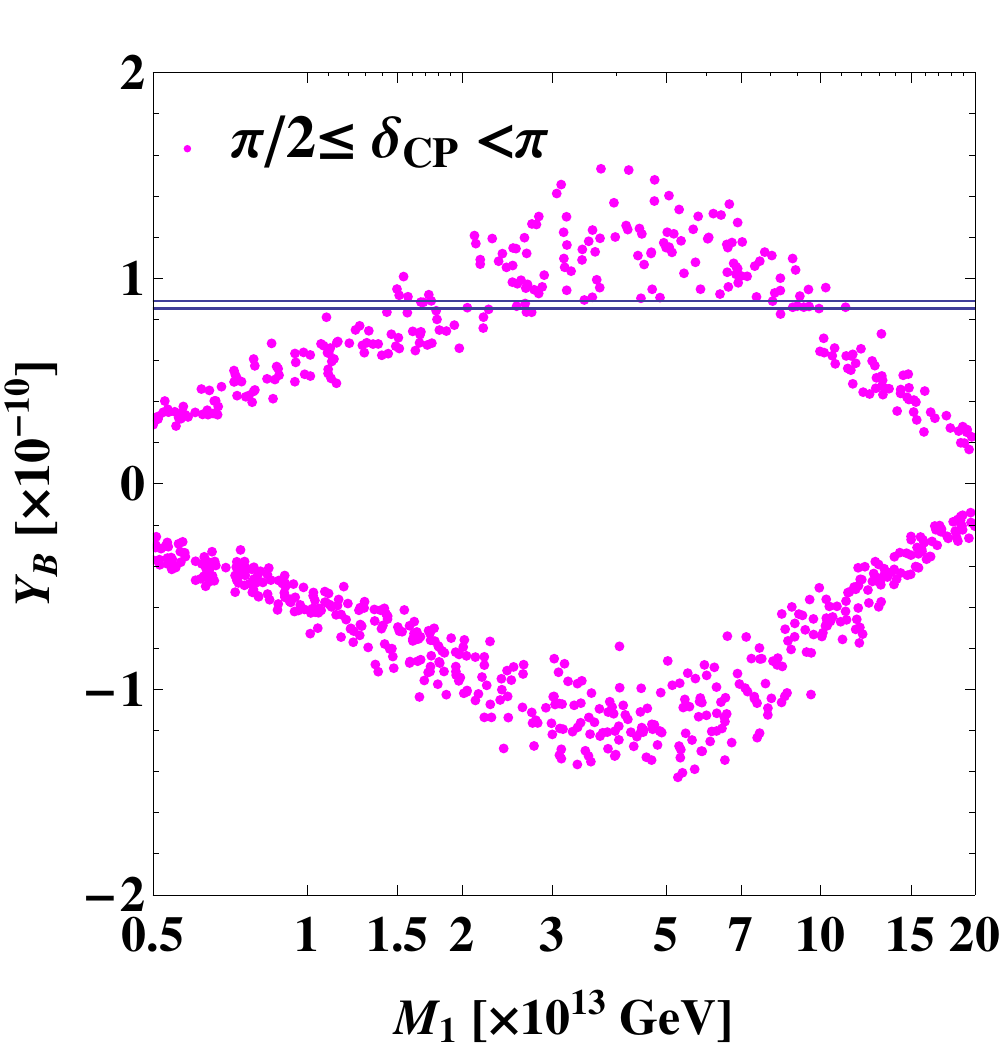}
    \vspace{1cm}
    \includegraphics[height=6cm]{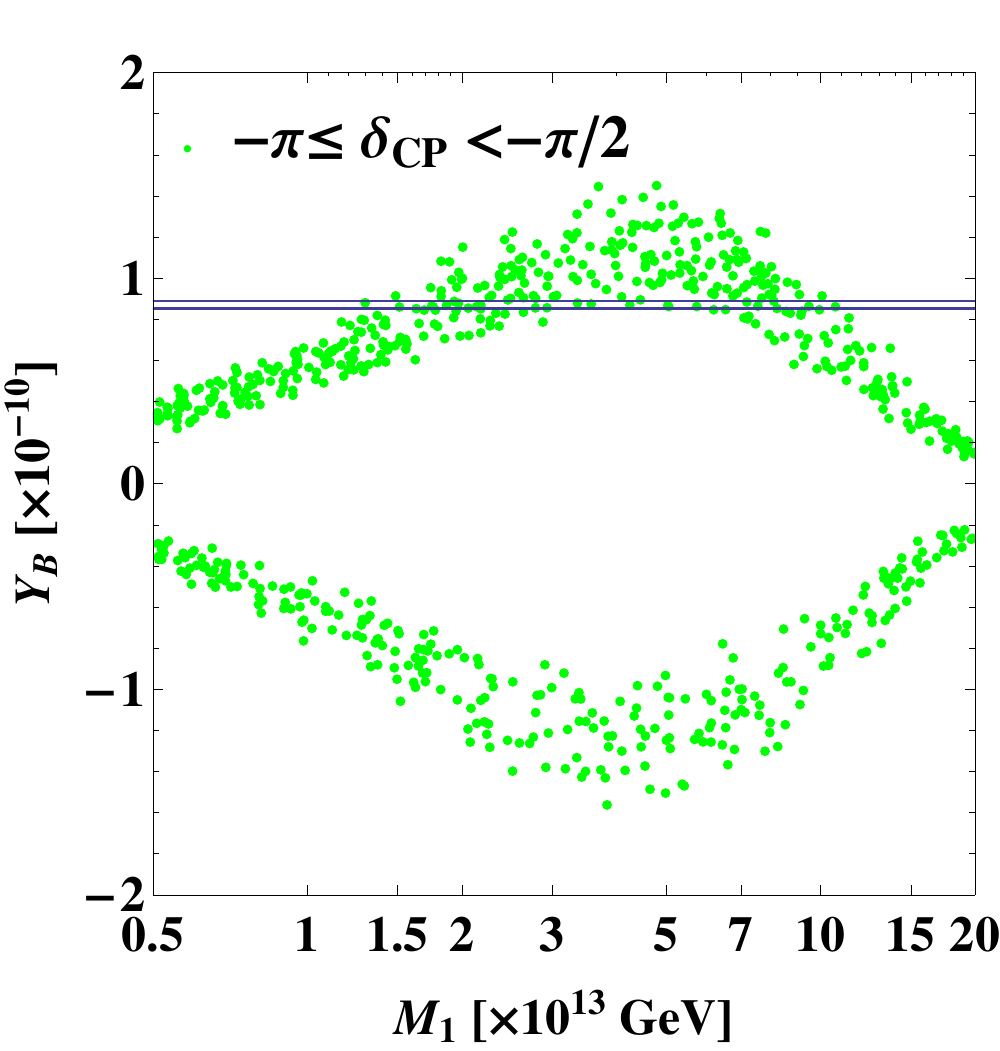}
    \hspace{1cm}
    \includegraphics[height=6cm]{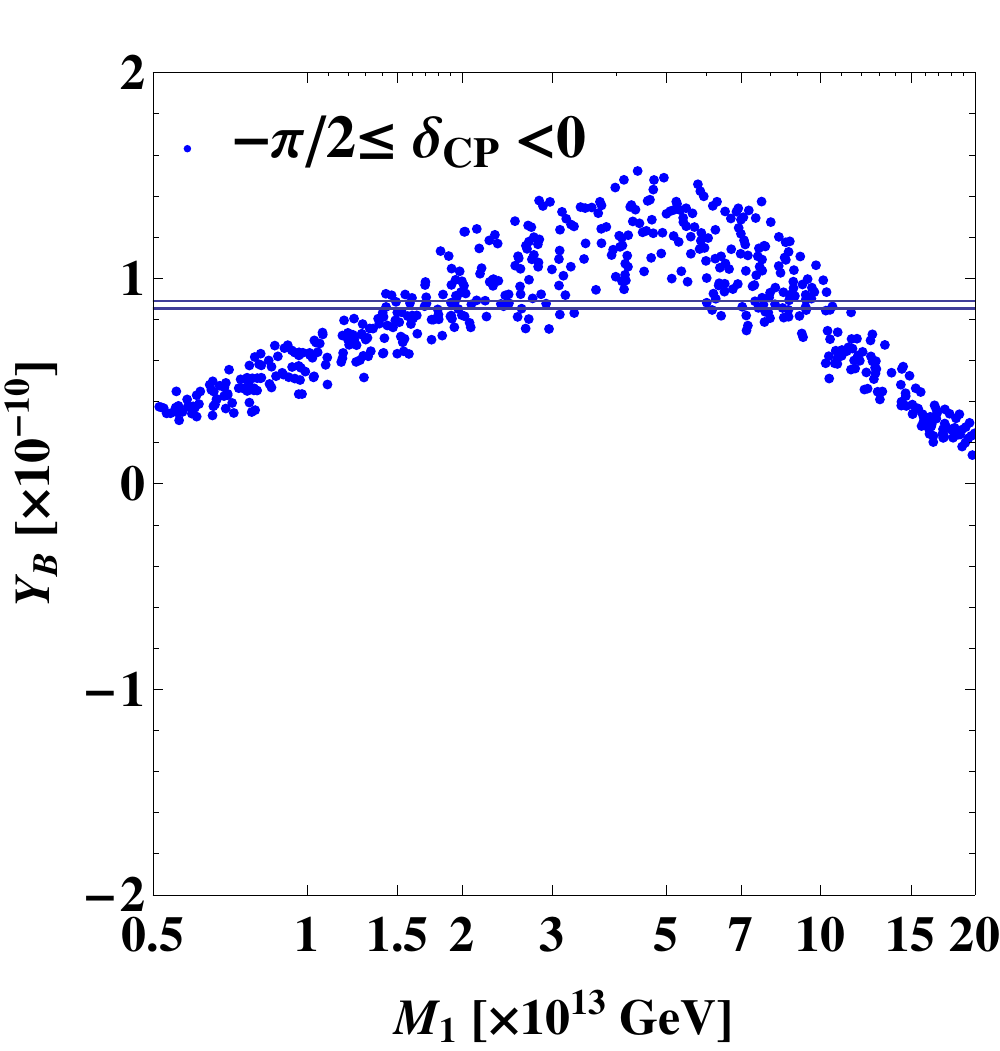}
  \caption{
	The yields of the BAU in terms of the mass of the lightest right-handed neutrino
	for different values of $\delta_{\rm CP}$.
	$\delta_{\rm CP}$ is taken in the range
	$[0, \pi/2]$ (top-left), $[\pi/2, \pi]$ (top-right),
	$[-\pi, - \pi/2]$ (bottom-left) and $[-\pi/2, 0]$ (bottom-right).
	The solid lines are upper or lower bound for the observational value of $Y_B$.
	}
  \label{M1vsYB}
  \end{center}
\end{figure}

Next, we discuss the magnitude of the BAU yield.
We find that the yield  can be at most the same order of the observed value of the BAU (1).
This is because the model predicts a relatively large value of
the effective neutrino mass of the leptogenesis $\tilde m_1$ which is defined as $\tilde m_1=(Y_{\nu} {Y_{\nu}}^{\dagger})_{11}{v_u}^2/M_1$. We find numerically $\tilde m_1 \simeq 54 - 57 \ \rm{meV}$, and then the strong wash-out effect is inevitable.
This leads to an important cevennsequence that the lightest right-handed neutrino 
should be in the mass range $M_1 \simeq (1.5 - 10) \times 10^{13}$~GeV.
As can be seen from Fig.~\ref{M1vsYB}, we find that the dependence of the magnitude of the BAU on the lightest right-handed neutrino mass changes at $M_1 \simeq 4.0 \times 10^{13}$~GeV. At $M_1 \lesssim 4.0 \times 10^{13}$~GeV, the larger $M_1$ is, the larger the magnitude of the generated BAU is. On the other hand at $M_1 \gtrsim 4.0 \times 10^{13}$~GeV, 
the larger $M_1$ is , the smaller the magnitude of the generated BAU is. This is because the larger $M_1$ is, the more the wash-out effect of the $\Delta L=2$ processes is important. Thereby, the lightest right-handed neutrino mass is restricted to the specific range ($M_1 \simeq (1.5 - 10) \times 10^{13}$~GeV) in order to explain the observed BAU.
%%%%%%%%%%%%%%%%%%%%%%%%%%%%%%%%%%%%%%%%%%%%%%%%%%%%%%%%%%%%%%%%%%%%%% 
%%%%% ** Figure ** %%%%%%%%%%%%%%%%%%%%%%%%%%%%%%%%%%%%%%%%%%%%%%%%%%%
%%%%%%%%%%%%%%%%%%%%%%%%%%%%%%%%%%%%%%%%%%%%%%%%%%%%%%%%%%%%%%%%%%%%%%

%%%%%%%%%%%%%%%%%%%%%%%%%%%
\section{Conclusions}
We have considered the leptogenesis in the model with three right-handed neutrinos
introducing the modular $A_4$ invariance.  
The model is very predictive in the sense that 
all the parameters apart from the overall scale of right-handed neutrino masses are determined within the limited ranges
in order to be consistent with the observed values of charged lepton masses
as well as mixing angles and masses of active neutrinos.
We have shown that the observed value of the BAU can be explained 
when the mass of the lightest right-handed neutrino is 
$(1.5 - 10)\times 10^{13}$~GeV.  This means that 
the successful baryogenesis determines the absolute masses 
of all right-handed neutrinos.

We have also shown that the sign of the BAU is strongly related with the CP violating parameters,
Majorana and Dirac phases, since the possible breaking pattern of the CP symmetry 
is very limited in the considering model.
In fact, the positive sign of the BAU is realized only for the unique range of 
Majorana phases, namely $\alpha_{21} \sim 1.3~\pi$ and $\alpha_{31} \sim 1.5~\pi$. Moreover, we have shown that  
the precise measurements of $\sin^2 \theta_{23}$
and $\delta_{\rm CP}$ provide a crucial test for the correct sign of the BAU 
in the considering baryogenesis scenario.

%%%%%%%%%%%%%%%%%%%%%%%%%%%%%%%%%%%%%
%%%%% acknowledgement %%%%%
\vspace{0.5cm}
\noindent

{\bf Acknowledgement} 

We would like to thank K.~Takagi and M.~Tanimoto for useful discussions.
This work is supported by 
JSPS KAKENHI Grant Numbers 17K05410, 18H03708, and 19H05097 (TA) and 
JSPS Grants-in-Aid for Scientific Research 18J11233 (THT).
TA, YH and TY thank the Yukawa Institute for Theoretical Physics at Kyoto University for the useful discussions during "The 47th Hokuriku Spring School" (YITP-S-19-01).
%%%%%%%%%%%%%%%%%%%%%%%%%%%%%%%%%%%%%
%%%%%%%%%%%%%%%%%%%%%%%%%%%%%%%%%%%%%%%%%%%%%%%%%%%%%%%%%%%
%%%%%%%%%%%%%%%%%%%%%%%%%%%%%%%%%%%%%%%%%%%%%%%%%%%%%%%%%%%
%%%%%%%%%%%%%%%%%%%%%%%%%%%%%%%%%%%%%%%%%%%%%%%%%%%%%%%%%%%%%%%%%%%%%%%%

%%%%%%%%%%%%%%%%%%%%%%%%%%%%%%%%%%%%%%%%%%%%%%%%%%%%%%%%%%%

%%%%%%%%%%%%%%%%%%%%%%%%%%%%%%%%%%%%%%%%%%%%%%%%%%%%%%%%%%%%%%%%%%%%%%%%  
%%%%%%%%%%%%%%%%%%%%%%%%%%%%%%%%%%%%%%%%%%%%%%%%%%%%%%%%%%%%%%%%%%%%%%%%%%%%%%
%%%%%%%%%%%%%%%%%%%%%    References    %%%%%%%%%%%%%%%%%%%%%%%%%%%%%%%%%%%%%%%
%%%%%%%%%%%%%%%%%%%%%%%%%%%%%%%%%%%%%%%%%%%%%%%%%%%%%%%%%%%%%%%%%%%%%%%%%%%%%%

\end{document}